\documentclass[sigconf]{acmart}

\AtBeginDocument{%
  }

\copyrightyear{2026}
\acmYear{2026}
\setcopyright{cc}
\setcctype{by}
\acmConference[CHI '26]{Proceedings of the 2026 CHI Conference on Human Factors in Computing Systems}{April 13--17, 2026}{Barcelona, Spain}
\acmBooktitle{Proceedings of the 2026 CHI Conference on Human Factors in Computing Systems (CHI '26), April 13--17, 2026, Barcelona, Spain}
\acmPrice{}
\acmDOI{10.1145/3772318.3791001}
\acmISBN{979-8-4007-2278-3/2026/04}

\usepackage{array}       
\usepackage{natbib}
\usepackage{booktabs}    
\usepackage{caption}

\usepackage{booktabs}   
\usepackage{graphicx}

\usepackage{listings}
\usepackage{xcolor}

\lstdefinestyle{promptstyle}{
    basicstyle=\ttfamily\small,   
    breaklines=true,              
    columns=fullflexible,         
    showstringspaces=false,       
    frame=single,                 
    framerule=0.5pt,              
    xleftmargin=0.5em,            
    framexleftmargin=0.5em,       
    numbers=left,            
    numbersep=0pt,           
    numberstyle=\tiny\color{gray}, 
}

\begin{document}

\title[Tracing Everyday AI Literacy Discussions at Scale]{Tracing Everyday AI Literacy Discussions at Scale: How Online Creative Communities Make Sense of Generative AI}

\author{Haidan Liu}
\affiliation{%
  \institution{Simon Fraser University}
  \city{Burnaby}
  \country{Canada}}
\email{haidanl@sfu.ca}

\author{Poorvi Bhatia}
\affiliation{%
  \institution{Simon Fraser University}
  \city{Burnaby}
  \country{Canada}}
\email{poorvi\_bhatia@sfu.ca}

\author{Nicholas Vincent}
\affiliation{%
  \institution{Simon Fraser University}
  \city{Burnaby}
  \country{Canada}}
\email{nvincent@sfu.ca}

\author{Parmit K Chilana}
\affiliation{%
  \institution{Simon Fraser University}
  \city{Burnaby}
  \country{Canada}}
\email{pchilana@sfu.ca}

\renewcommand{\shortauthors}{Liu et al.}

\begin{abstract}

Developing AI literacy is increasingly urgent as generative AI reshapes creative practice. Yet most AI literacy frameworks are top-down and expert-driven, overlooking how literacy emerges organically in creative communities. To address this gap, we performed a large-scale analysis of 122k Reddit conversations from 80 creative-oriented subreddits over a time period of three years. Our analysis identified four consistent themes in AI literacy-related discussions, and we further traced how discourse shifted alongside major AI events.  Surprisingly, creators primarily frame AI literacy around how to use tools effectively—foregrounding practice and task skills—while discussions of AI capabilities and ethics surge only around high-profile events. Our findings suggest that AI literacy is dynamic, practice-driven, and event-responsive rather than static or purely conceptual. This study provides insights for researchers, designers, and policymakers to develop learning resources, community support, and policies that better promote AI literacy in creative communities.

\end{abstract}




\maketitle

\section{Introduction}

Generative AI now enables individuals without prior professional experience to engage in creative tasks such as image generation, video production, and storytelling that once required years of specialized training or deep domain expertise \cite{sachit, federico}. Today, creators can generate sophisticated visual content in seconds by simply typing a natural language prompt. While this shift lowers traditional barriers to creation, it has surfaced a new kind of digital divide: not one based on access to AI tools, but on the knowledge and skills required to use them effectively \cite{aiprof}. Despite the widespread adoption of AI, most users remain ill-equipped to move beyond surface-level use, lacking the literacy to meaningfully and critically engage with these systems \cite{static, Burgsteiner, ghallab}.

AI literacy has recently been defined as: ``the knowledge required to understand, the operative skills to engage with, the awareness needed to critically evaluate AI technologies, including their ethical, societal, and practical implications'' \cite{biagini}. Building on earlier frameworks of media literacy, which emphasized critical consumption of mass media \cite{medialiteracy}, and data literacy, which centered on ethical interpretation and use of data \cite{javier}, AI literacy introduces distinctive challenges. Unlike traditional media or data sources, which exist as static objects to be passively consumed, current AI systems require prompting and querying, involve active co-creation with users, and demand dynamic interaction skills. Additionally, users may have to continuously adapt their strategies to accommodate frequent model updates, as techniques effective with one AI model might prove inadequate with the next \cite{gagan, aiwildcard, oppenlaender2024prompting}. This perpetual adaptation cycle makes achieving and maintaining AI literacy a particularly complex and ongoing challenge.

\begin{table*}[t]
\centering
\caption{Summary of analysis stages. Topic modeling and qualitative coding were used to identify and consolidate themes for RQ1, while classification and temporal analysis enabled us to examine how AI literacy has evolved for RQ2.}
\label{tab:analysis-pipeline}
\small
\renewcommand{\arraystretch}{1.2} 
\begin{tabular}{p{2.95cm} p{3.8cm} p{8.2cm}}
\toprule
\textbf{Stage} & \textbf{Data size / sample} & \textbf{Description} \\
\midrule
Data collection & 122,506 posts; 1,554,368 comments & Reddit discussions extracted through keyword filtering and after preprocessing \\

Topic modeling & Full dataset collected from data collection stage & Generated initial topics discovered \\

Theme consolidation & 6 themes merged from topic modeling results & Grouped related topics into broader themes, guided by Long and Magerko's AI literacy definition \\

Grounded qualitative analysis and theme refinement & 900 sampled conversations & Conducted in-depth coding, resulting in 8 themes used for classification \\

Content classification & 122,506 conversations & Labeled all conversations and filtered out those labeled as unrelated content, resulting in 112,735 conversations \\

Temporal analysis & 112,735 conversations & Tracked how 7 AI-related themes have shifted to major AI events and focused on reporting 4 literacy-related themes \\
\bottomrule
\end{tabular}
\end{table*}

Researchers in HCI and Education have proposed a range of AI literacy frameworks to guide educational and design interventions \cite{heyder,durilong20, understand1, druga2019, ng,ngb}, often by interviewing experts \cite{kamila2023} or reviewing existing literature \cite{ng, durilong20}. While expert-driven approaches highlight what individuals \textit{should} know to be AI literate, they provide limited insight into \textit{how} everyday creators actually encounter and make sense of AI in practice \cite{xie2025, aipainting}. Yet fostering AI literacy among non-AI experts---especially creators who use AI tools every day---is crucial \cite{ng, ngb}. To complement expert perspectives with a more grounded view of how AI is experienced in practice, researchers have started exploring informal and social learning environments, such as \textit{Reddit} \cite{aipainting}, but this work has been limited to short-term analyses of single subreddits on specific topics, such as AI painting.

Building on this perspective, we extend current understandings of AI literacy by analyzing how people discuss and learn about AI within everyday online creative communities. We use the term \textit{creators} to refer primarily, though not exclusively, to individuals who produce visual media. Because AI literacy develops gradually through repeated use, problem-solving, and reflection, studying it requires sustained, longitudinal evidence. In this study, we analyze discussions across 80 creative-related subreddits over three years, using data obtained through Reddit’s official \textit{Reddit for Researchers} program. We curate a large-scale subset of AI-related conversations through targeted keyword queries and adopt a grassroots perspective to trace how major technological developments shape community dialogue over time. 
Our study is guided by two research questions: 


\begin{itemize}
     \item RQ1: What topics related to AI literacy are discussed in online creative communities?
     \item RQ2: How have AI-related discussions in online creative communities evolved over time around major AI events (e.g., model release, tools launch or policy changes)?
\end{itemize}

We summarized the scale and scope of each stage in our analysis pipeline in Table \ref{tab:analysis-pipeline}. To uncover the kinds of AI literacy themes that emerge in creative communities (RQ1), we first applied computational topic modeling to our corpus of 122,506 Reddit posts and 1,554,368 comments extracted using keyword-based filtering. The initial results revealed a variety of topics ranging from AI's social impact to practical tool-oriented help-seeking behavior. To interpret how these themes relate to AI literacy in creative practice, we drew inspiration from Long and Magerko's~\cite{durilong20} definition and used it as a guiding lens to situate each theme with respect to one or more of the three core components of AI literacy. To better capture the complexity of real-world AI discourse, we also conducted an in-depth qualitative analysis by manually examining 900 sampled conversations. Together, these analyses revealed AI literacy competencies not captured in existing frameworks---such as workflow integration practices and output quality assessment---highlighting how creative communities can complement our understanding of AI literacy.

To investigate how AI literacy has evolved (RQ2), we classified 122,506 conversations from April 2022 to February 2025 and conducted a temporal analysis of 112,735 (excluding those labeled as not-related in the content classification step, e.g., commissions). We traced discussion patterns across major AI events, from model releases to policy shifts, examining both immediate reactions and sustained changes. This temporal analysis allowed us to move beyond a static view of AI literacy and instead capture its dynamic, event-driven nature within online creative communities.

Our findings reveal that creators in online communities prioritize the \textit{practical} dimensions of AI literacy, generally focusing on how to get work done rather than understanding how AI works internally. This contrasts with existing AI literacy frameworks, which often emphasize conceptual understanding and foundational knowledge as prerequisites for being AI literate \cite{ng, kandlhofer2016, druga2019, druga2022, wong_broadening_2020}. This disconnect suggests that efforts to promote AI literacy for creators may benefit from foregrounding applied skills and situated training over abstract knowledge. By highlighting this, our study points to opportunities for interventions that better support creators' real-world learning needs.

To summarize, our work contributes to HCI research in the following ways. First, we offer a bottom-up, practice-grounded account of AI literacy that contrasts with existing top-down, expert-driven AI literacy frameworks (e.g., \cite{kamila2023}); by deriving themes directly from creators' everyday discussions, we surface literacy practices, such as workflow integration, collaborative troubleshooting, capability probing, and navigating ethics around major AI releases, that do not appear in expert frameworks or in single-community studies. Second, our mixed methods design combines large scale topic modeling with inductive qualitative analysis to provide a more complete and nuanced picture of how AI literacy unfolds in creative communities, complementing prior work that examines AI literacy through either a qualitative lens (e.g., \cite{ng, biagini, kamila2023}) or a quantitative one (e.g., \cite{aipainting}). Our three-year dataset of 122K conversations across creative communities (e.g., visual art, writing, design) enables us to trace how discourse evolves alongside model releases, tool updates, and platform-level disruptions. Finally, we show how AI adoption is fostering new forms of peer-to-peer learning and community knowledge---building within social platforms at a moment when traditional Q\&A infrastructures (e.g., \textit{StackOverflow}) are in decline \cite{delrio, samirakabir}. Together, these contributions advance our understanding of AI literacy as a socially situated, evolving practice, and inform the design of systems that better support individuals and communities in developing AI literacy.

\section{Related Work}

To situate our insights on AI literacy within online communities, we draw on three strands of research: (1) frameworks for understanding AI literacy, (2) studies of art and creative education, and (3) research on online communities as informal learning spaces.

\subsection{AI Literacy: Definition and Dimensions}

AI literacy lacks a single agreed-upon definition, with ongoing debates about its scope, dimensions, and pedagogical emphasis \cite{Burgsteiner}. Most accounts converge on a set of core competencies that span conceptual understanding, tool use, and ethical awareness. \citet{durilong20}, for example, define AI literacy as ``a set of competencies that enables individuals to \textit{critically evaluate AI technologies; communicate and collaborate effectively with AI; and use AI as a tool} online, at home, and in the workplace''. Importantly, they argue that digital and computational literacy are not prerequisites.

Subsequent frameworks have elaborated this foundation by identifying multiple dimensions of literacy. Heyder and Posegga~\cite{heyder}, for example, distinguish between \textit{functional} (understanding how AI works), \textit{critical} (evaluating and questioning AI), and \textit{sociocultural} (considering norms, organizational cultures, and adoption practices) literacies. Similar multidimensional frameworks echo across contexts, emphasizing competencies such as understanding AI concepts \cite{ng, ngb, kandlhofer2016, kim2021, druga2019}, applying AI tools in practice \cite{druga2019, druga2022, wong2020, kim2021, laupichler2022}, evaluating AI outputs \cite{durilong20, wong2020, ng,ngb}, and addressing ethics and societal concerns \cite{wong2020, ng, ngb}. Several also highlight data, including recognizing algorithmic bias \cite{unicef}, examining data collection \cite{hermann2022}, or treating data literacy as a component of AI literacy \cite{unesco}. Ng et al.~\cite{ng} consolidate these into four aspects: knowing and understanding AI, using and applying AI, evaluating and creating with AI, and addressing ethical issues.

While these AI frameworks provide valuable conceptual clarity, they are largely top-down and expert-driven, focusing on what people \textit{should} know. Far less is understood about how AI literacy actually emerges \textit{in practice}, particularly in informal, community-driven settings. Our study addresses this gap by analyzing Reddit everyday discussions across multiple creative subreddits, revealing AI literacy as a bottom-up, socially situated process involving workflow integration, help-seeking and troubleshooting, capability awareness and exploration, and ethical sense-making around major AI events.

\subsection{How Creatives Learn About AI in Formal Education}

In formal education, researchers have explored contrasting approaches to teaching AI to creatives. One perspective argues that practitioners require a basic technical understanding of machine learning \cite{king2017, hebron2016} through books, tutorials, or courses \cite{king2017, hebron2016}. Another perspective contends that creatives can engage productively without mastering the underlying mathematics, instead focusing on abstractions of AI's capabilities and illustrative applications to envision how these technologies could support practice \cite{yang2018}.

Several pedagogical approaches attempt to bridge these views. Fiebrink~\cite{Fiebrink2019} demonstrates how interactive tools foster functional understanding through hands-on experimentation rather than abstract study. Building on this, Huang et al.~\cite{huang_experiential_2023} propose a vision-first pedagogy that centers creative goals and aesthetic values, enabling learners to explore how AI can serve their personal visions. This approach produced diverse sensual, conceptual, and discursive outcomes, and promoted deeper engagement than conventional instruction. Yet challenges remain: Flechtner and Stankowsk~\cite{aiwildcard} observe that AI education in design remains fragmented---workshops inspire but fade, technical courses alienate, and studio projects often treat AI as a ``wildcard''. They advocate institutional AI labs that offer long-term resources, mentorship, and interdisciplinary support to foster more durable forms of literacy.

Taken together, this body of work highlights both the promise and the limitations of formal educational strategies for cultivating AI literacy. Our study extends this literature by examining Reddit as an informal site where creators develop literacy through everyday participation, capturing grassroots practices beyond structured pedagogy.

\subsection{Online Communities as Informal Learning Spaces}

Online communities such as Reddit, Discord, and dedicated forums often function as informal learning hubs where members collectively explore new creative technologies \cite{Kim2016Mosaic}. Interest-driven communities encourage the sharing of artifacts and techniques, fostering creativity and peer collaboration without formal instruction \cite{CoRemix}. Knowledge is curated as members post work, exchange resources, and engage in ongoing discussion with peers that scaffold community-generated learning \cite {cheng_how_2022}.

For creators, these online forums have long provided crucial avenues for skill development outside formal education \cite{scratch}. With the rise of generative AI, prompting itself has emerged as a creative skill in its own right \cite{oppenlaender2024prompting}. Prior work has also shown that visual artists view both prompts and prompt templates as part of the artwork itself \cite{chang2023promptartists}. Studies of \textit{Midjourney}'s Discord community, for instance, revealed how creators approached prompting as a craft developed through trial, error, and collective knowledge-sharing \cite{oppenlaender2022}. Within these spaces, members adopt distinct roles---innovators, porters, conservators, service providers, and practitioners---to sustain ecosystems around AI art \cite{oppenlaender2022}.  

\begin{figure*}[t]
  \centering
  \includegraphics[width=1\linewidth]{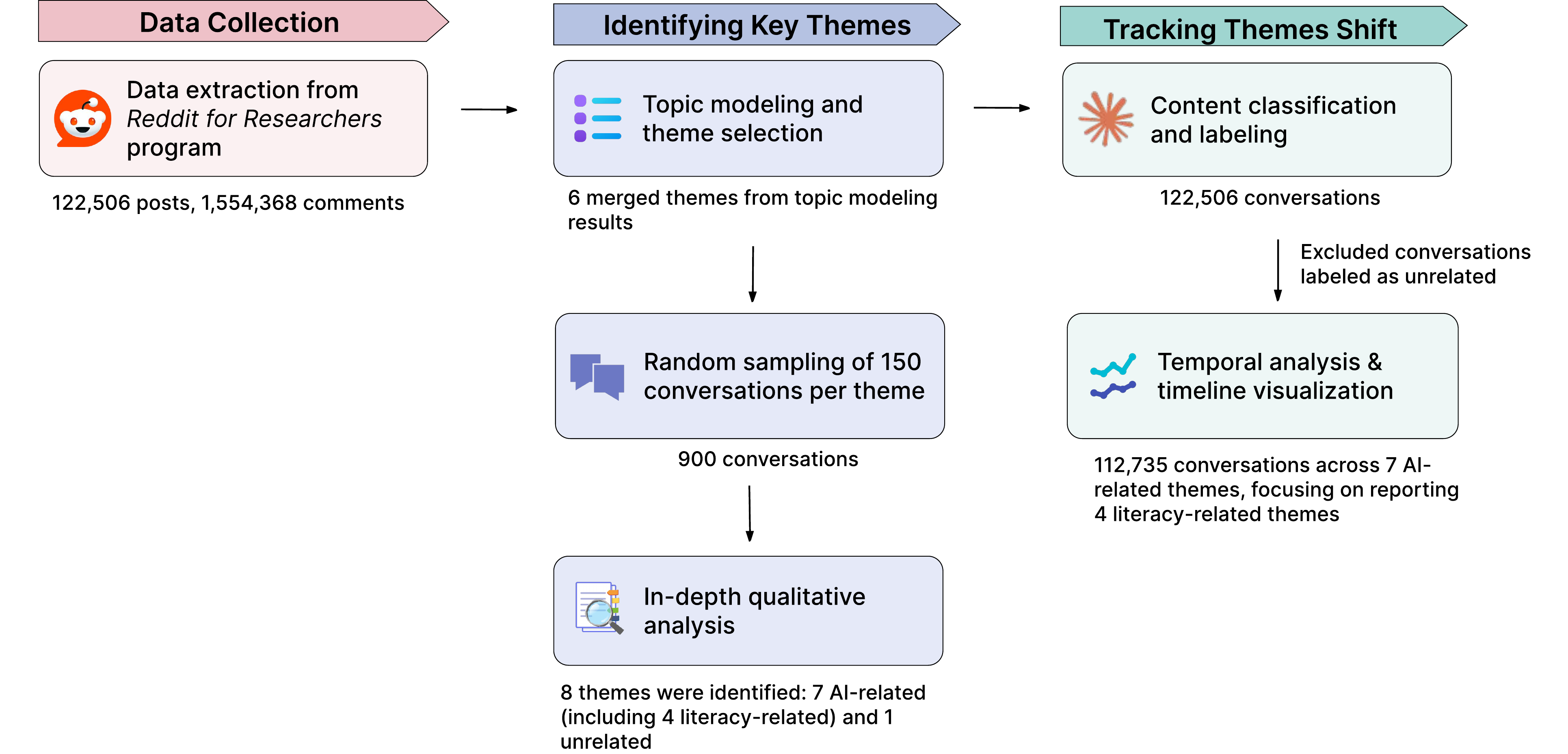}
  \caption{Overview of our mixed-methods pipeline. We collected 122k Reddit posts and 1.5m comments, identified 6 initial themes after merging topic modeling results on posts, and then conducted qualitative analysis on 900 conversations. Through this analysis, we refined the themes and arrived at 8 final themes (see Figure~\ref {fig:process}), including 7 AI-related themes and 1 unrelated. We then classified all conversations, conducted a temporal analysis of 7 AI-related theme dynamics over time, and focused on reporting 4 more literacy-related themes among them in the results section.}
  \label{fig:process}
  \Description{The figure illustrates a three stage research pipeline for analyzing AI literacy discussions on Reddit. The process begins with data collection, where posts and comments are extracted from Reddit using the Reddit for Researchers program, resulting in 122,506 posts and 1,554,368 comments. The second stage focuses on identifying key themes through topic modeling and theme selection. Six merged themes are derived from topic modeling results, followed by random sampling of 150 conversations per theme for a total of 900 conversations. These sampled conversations undergo in depth qualitative analysis, leading to the identification of eight themes, including seven AI related themes, four of which are literacy related, and one unrelated theme. The final stage tracks how themes shift over time through content classification and labeling of all conversations. After excluding unrelated conversations, 112,735 AI related conversations remain. These are used for temporal analysis and timeline visualization, with a focus on reporting trends across four AI literacy related themes.}
\end{figure*}

Reddit has likewise emerged as a site where creative communities negotiate practices and norms around AI. Recent works have identified central themes---such as model usage, ethics, and procedural guidance---showing how community norms and shared concerns begin to take shape and highlighted sustained engagement with experimentation, aesthetic reflection, prompt design, and ethical debates \cite{aipainting}. These studies suggest that how Reddit functions is not merely as a technical help forum, but as a dynamic environment where values, practices, and creative norms are continually shaped.

Building on this foundation, our work examines how participation in Reddit contributes to the development of AI literacy. Prior studies have documented how creators share knowledge, exchange resources, and experiment collectively within individual communities (e.g.,\cite{aipainting}). However, these accounts are typically limited to single subreddits, short observation periods, or a single methodological lens---either qualitative or quantitative. We extend this work through a large-scale, mixed-methods, three-year analysis across multiple creative subreddits, enabling us to identify how competencies emerge across communities over time and to conceptualize AI literacy as socially situated and continually evolving within creative online communities.

\begin{figure*}[t]
  \centering
  \includegraphics[width=1.0\linewidth]{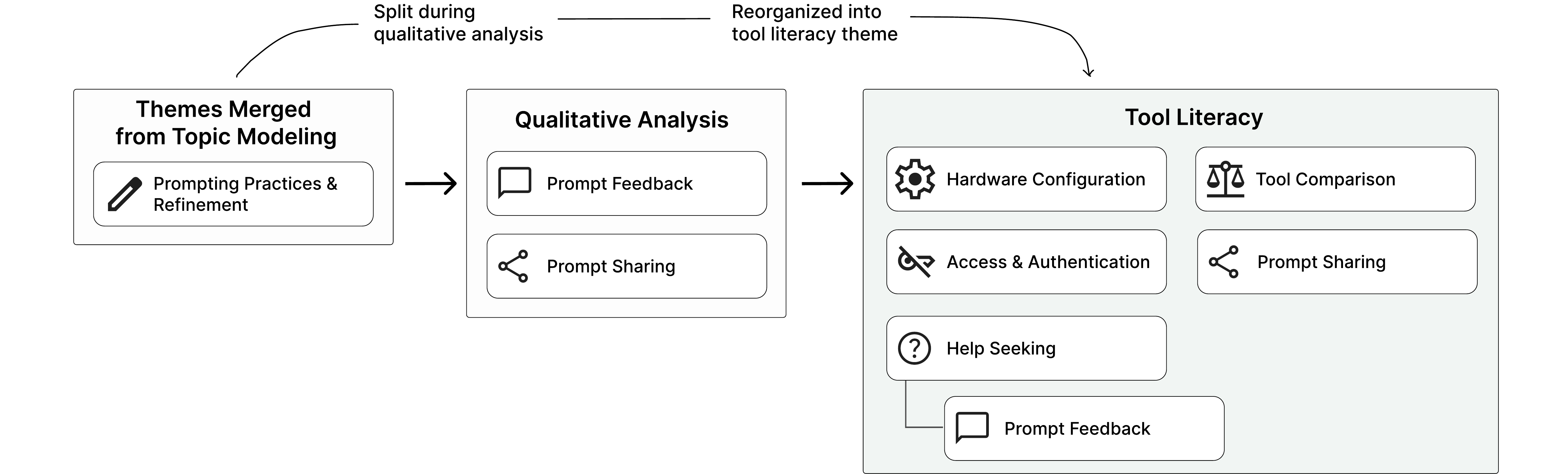}
  \caption{Overview of how the topic-modeled theme Prompting Practices \& Refinement was reorganized during qualitative analysis. The theme was divided into prompt sharing---when users post full prompts as examples of prompt crafting and prompt feedback---when users share prompts to seek suggestions or improve results. These subcategories were then folded into the broader Tool Literacy theme, with prompt feedback ultimately placed under the help-seeking subcategory.}
  \label{fig:thememergeprocess}
  \Description{The figure shows how a topic modeled theme was reorganized during qualitative analysis into a more structured AI literacy theme. On the left, a theme labeled Prompting Practices and Refinement is shown as the output of topic modeling. During qualitative analysis, this theme is split into two distinct subcategories: prompt sharing and prompt feedback. Prompt sharing refers to users posting full prompts to seek suggestions or improvements, while prompt feedback refers to discussions about evaluating, refining, or critiquing prompts. On the right, these subcategories are reorganized into the broader Tool Literacy theme. Prompt sharing is grouped alongside other tool related practices such as hardware configuration, access and authentication, tool comparison, and help seeking. Prompt feedback is ultimately placed under the help seeking subcategory.}
\end{figure*}

\section{Data and Methods}
We analyzed Reddit discussions through a multi-stage pipeline combining topic modeling, qualitative coding, content classification, and temporal analysis.

\subsection{Dataset}
Our dataset was obtained through Reddit's official \textit{Reddit for Researchers} program \cite{reddit4}, which provides governed and privacy-conscious access to public Reddit content. As a result, our dataset includes only posts and comments available through February 2025, reflecting the program's archival coverage at the time of data collection. Furthermore, following emerging best practices in taking a human-centered approach to large-scale analysis of social media, all quotes from Reddit have been lightly paraphrased for readability and to reduce searchability while maintaining semantic fidelity to the original post \cite{studyingreddit}.

To examine large-scale discourse on AI literacy in creative communities, we collected Reddit data from April 2022 to February 2025, beginning with the announcement of \textit{DALL·E 2} and a wave of generative AI tools launched that same year---Midjourney and \textit{ChatGPT}, which catalyzed widespread public engagement with generative AI tools \cite{towardsailiteracy}.

Our analysis required identifying where these conversations occurred and how they were framed. This involved two steps: (1) selecting relevant subreddits where visual creators may engage with generative AI (e.g., \textit{r/aiArt}), and (2) constructing a structured keyword taxonomy to filter and categorize AI-related discourse. 

We used a multi-stage process to identify relevant online communities. Starting with creative-focused subreddits cited in prior work on AI-generated imagery \cite{prevalence}, we expanded the list using online articles recommending subreddits for visual artists \cite{newartists, govisually, format, nyfa}. We also included subreddits dedicated to generative AI tools (e.g., Midjourney, ChatGPT) and broader AI-art communities \cite{aiartsubreddits, metapix}. To ensure relevance, we searched each subreddit for the term ``AI'' using Reddit's internal search. Subreddits were excluded if none of the top five posts referenced artificial intelligence meaningfully or if ``AI'' referred only to unrelated acronyms. This yielded a final set of 80 subreddits spanning both traditional art and AI-art communities (Appendix~\ref{appendix:subreddit}).

From these subreddits, we extracted posts and comments using a keyword-based filtering strategy. To capture the multidimensional nature of AI discourse, we developed a taxonomy with three categories: (1) generative AI tools, (2) AI-related concepts, and (3) image generation and manipulation techniques. Details and the complete list are provided in Appendix~\ref{appendix:keyword}.

\subsection{Study 1: Identifying Conversation Themes}

\subsubsection{LDA Topic Modeling and Theme Selection} For our first study, we aim to answer RQ1: What topics related to AI literacy are discussed in online creative communities? We began by extracting 129,016 posts and 1,595,663 comments containing AI-related keywords from the Reddit dataset. We removed non-linguistic artifacts and excluded non-English entries (using the Langdetect library) as well as posts containing five or fewer words. We deliberately retained stopwords at this stage as prior work suggests that keeping stopwords can improve topic interpretability \cite{stopword1,stopword2}. After preprocessing, 122,506 posts and 1,554,368 comments remained for analysis.

We applied Latent Dirichlet Allocation (LDA) \cite{blei} separately to the posts and comments using the Tomotopy library. After qualitatively reviewing, we selected topic counts \(K\) = 25 for posts and \(K\) = 20 for comments with \(\alpha = 0.1\) and \(\eta = 0.01\) and trained with 2,000 Gibbs-sampling iterations. After removing stopwords, we computed topic model coherence scores \cite{umass, cvmeasure, npmi} to guide model selection (see Table \ref{tab:comments_coherence_scores} and Table \ref{tab:posts_coherence_scores} in Appendix~\ref{appendix:topic_modeling}).
We excluded topics lacking semantic coherence or relevance to our research focus (e.g., jobs, photography). This resulted in 18 post topics and 16 comment topics (see Table \ref{tab:post_topics} and Table \ref{tab:comments_topic} in Appendix ~\ref{appendix:topic_modeling}). Since comment topics did not have extra special topics, we focused the subsequent analysis on post topics. 

We adopt Long and Magerko's~\cite{durilong20} definition of AI literacy as ``a set of competencies that enables individuals to critically evaluate AI technologies; communicate and collaborate effectively with AI; and use AI as a tool online, at home, and in the workplace''. Using these three components as a guilding lens, we merged related posts' topics into six themes. For example, we combined Prompt Design \& Image Styles (5.6\%) and Prompt Settings Techniques (1.1\%) into a broader theme, \textit{Prompting Practices \& Refinement (6.7\%)}, as both captured how creators iteratively refine prompts and settings to achieve desired outputs (reflecting the ``communicate and collaborate with AI'' and ``use AI as a tool'' components from Long and Magerko).

\subsubsection{Qualitative Conversations Analysis}

Prior research shows that although topic models surface frequently used terms, interpreting their meaning and relevance often requires deeper qualitative analysis \cite{gencoglu_machine_2023}. To complement the computational findings and better understand how topics were embedded in the conversational flow, we conducted a qualitative content analysis on a random sample of 900 conversations (150 per theme). Each conversation included the original post and up to the top five comments (sorted by score), totaling 3,439 items.

The topic modeling results served as a starting point to structure our initial codebook. We used top keywords and representative posts and comments from the topic modeling results to define an initial set of codes. While we drew on Long and Magerko's~\cite {durilong20} AI literacy definition as a reference when interpreting broad topic-model clusters, we did not use it as a fixed coding scheme. We employed an inductive analysis approach \cite{strauss1994grounded}, and our coding process was iterative: we continually reorganized, split, or merged codes to capture distinctions not visible in the topic model and to more accurately reflect emerging patterns in the data. For example, Sharing, Feedback \& Community was split into Community Engagement and AI Output Sharing to better reflect the range of social and creative sharing behaviors. Similarly, Prompting Practices \& Refinement was subsumed under Tool Literacy, as creators’ prompting strategies were understood as part of learning to use the tool effectively. We then divided these discussions into two subcategories---prompt sharing and prompt feedback---with the latter ultimately moved under the help-seeking subcategory in Tool Literacy (see Fig \ref{fig:thememergeprocess}). In some cases, we also introduced finer-grained subcodes to better capture variation within themes. For example, the AI tools general help-seeking theme produced a broader initial code labeled ``help-seeking'', which covered a wide range of support requests. During the process, the help-seeking category was further broken down into subcodes such as ``procedure help'' and ``troubleshooting''. In the Results section, we report the distribution and content of these help-seeking subcategories in more detail. 

A pilot set of 200 conversations was used to calibrate coding consistency. Two researchers first independently coded the pilot set and then compared their results to refine and revise the codes. Inter-rater reliability was assessed using Cohen's Kappa ($\kappa$) \cite{kappa}, with regular meetings held with a third researcher to resolve discrepancies, refine code definitions, and update labels. This process continued until a satisfactory $\kappa$ score of $\geq 0.80$ was achieved. After the initial codebook was established, we applied it to the remaining discussion conversations. Following a semi-open coding approach, the researchers inductively introduced new codes as novel themes emerged that were not captured in the initial set, continuing until no additional themes were identified. The complete codebook is provided in the Appendix~\ref{appendix:codebook}.

\subsection{Study 2: Tracking How Themes Shift Over Time}

\subsubsection{Content Classification and Labeling}

To classify conversations at scale, we tested rule-based methods, classical and deep learning models, and large language models (LLMs) prompting. Full details are provided in Appendix~\ref{appendix:classification_methods}. Evaluation was conducted on the 900 manually coded posts from our qualitative analysis, with a 600/300 train-test split.
 
We also tested how input formatting affected LLM performance. Including bot comments---automated replies from the explicitly identified as bots---raised accuracy from 56\% to 81\% and macro F1 from 34\% to 77\%. Though automated, these comments often contain topic-specific language that aligns closely with the original post. For example, comments like ``\textit{We kindly ask to respond to this comment with the prompt they used to generate the output in this post...}'' often appeared under AI-generated image posts, helping the model identify the ``AI Output Sharing'' category.

Based on these experiments, we selected Claude Sonnet 3.5 with full conversations (including bot comments) and few-shot prompting using 5 dynamically retrieved examples as our classification pipeline. This configuration achieved 81\% accuracy and a macro F1-score of 77\%. Performance was highest for ``Tool Literacy'' and ``Ethics and Responsible Use'', and lowest in ``Community Engagement'' and ``AI Tech Dynamics Sharing'' (F1-score = 67\%). The prompt is provided in Appendix~\ref{appendix:prompt}.

Using this pipeline, we analyzed 122,506 conversations into one of eight themes identified from our qualitative analysis (see Figure ~\ref{fig:process}).

\begin{figure}[tb]
  \centering
  \includegraphics[width=1.0\linewidth]{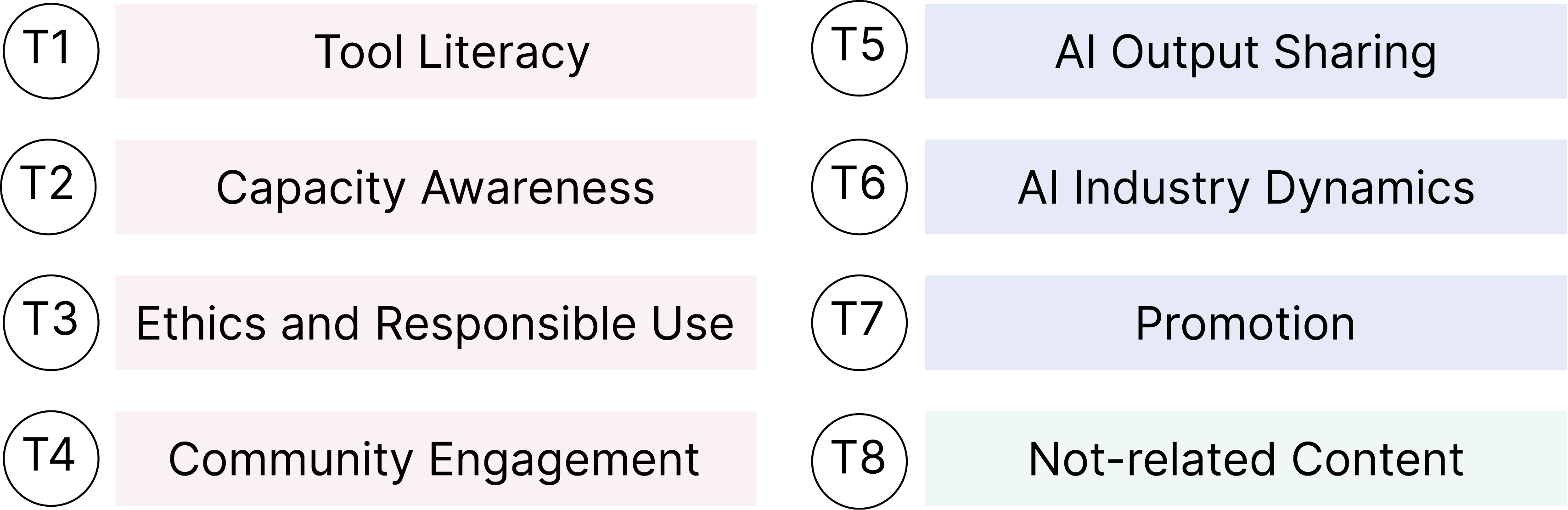}
  \caption{We identified 8 themes through qualitative analysis. During content classification, each conversation was labeled with one of these 8 themes. Among them, T1–T7 represent AI-related content, while T8 (Not-related Content) was excluded from the temporal analysis. In the results section, we focus on reporting themes T1–T4, as these themes are more directly connected to AI literacy.}
  \label{fig:process}
  \Description{The figure presents the eight themes identified through qualitative analysis, labeled T1 through T8. T1 is Tool Literacy, T2 is Capacity Awareness, T3 is Ethics and Responsible Use, and T4 is Community Engagement. These four themes represent core dimensions of AI literacy. T5 is AI Output Sharing, T6 is AI Industry Dynamics, and T7 is Promotion, which capture AI related discussions that are not directly focused on literacy. T8 represents Not related Content and includes conversations unrelated to AI. The figure indicates that during content classification, each conversation was assigned to one of the eight themes. For temporal analysis, only AI related themes T1 through T7 were included, while T8 was excluded. The results section further focuses on themes T1 through T4 because they are most directly connected to AI literacy.}
  
\end{figure}

\subsubsection{Temporal Theme Evolution and Event Analysis}

To construct a list of major AI events for our analysis, we examined the discussion corpus using a predefined list of AI tools. We then aligned these tools with external events identified through news coverage, such as public availability of tools and feature launches. The list also includes events that triggered public discourse (e.g., the deepfake controversy involving ElevenLabs \cite{elevenlabs1, elevenlabs2}) and significant platform-level changes, such as Reddit's API pricing shift \cite{protest1, protest2, protest3}.

We aggregated conversations by month, treating multiple AI events within the same month as a single analytical unit. For the temporal analysis, we excluded ``Not-related content'' and focused on the remaining 112,735 AI-related conversations, including 112,735 posts and 365,981 comments. We count activity for a window before and after the event: using the month immediately preceding each event as the universal baseline period. We then measured thematic shifts in the one-, two-, and three-month periods following the event month. Our main findings focus on reporting the one-month window to capture immediate shifts in discussions around AI events. For each theme, we calculated two types of change: (1) proportion change, measured in percentage points as the change in a theme's share of total AI-related conversations; and (2) relative change, the percentage increase or decrease in conversation volume relative to the baseline month. 

\begin{table*}[tb]
\centering
\caption{Overview of the six merged themes from our topic modeling analysis of posts. Each theme is shown with its relative share of posts and a description of the kinds of posts it captures. Tool-Related Complaints and Basic Setup \& Getting Started Help were the most common, together accounting for nearly half of all posts.}
\label{tab:theme_differences}
\small
\setlength\tabcolsep{6pt}
\renewcommand{\arraystretch}{1.3}
\begin{tabular}{@{} p{4cm} p{1.5cm} p{8.5cm} @{}} 
\toprule
\textbf{Theme} & \textbf{Percentage} & \textbf{Description} \\
\midrule

Tool Related Complaints & 26.2\% &
Captures frustrations when tools fail, outputs are poor, or policies feel restrictive. Posts describe crashes, errors, degraded quality, or complaints about terms of service.  \\

Basic Setup \& Getting Started Help & 23.8\% &
Covers entry-level posts where users seek basic guidance, often framed as requests for ``help'' (e.g., ``can someone help...'', ``how do I...''). These include troubleshooting access issues, asking about hardware or software compatibility, clarifying tool features, or requesting model and tool recommendations. \\

Model Training \& Workflow Customization & 13.3\% &
Involves advanced practices such as fine-tuning models (e.g., \textit{LoRA}, \textit{DreamBooth}), configuring \textit{ComfyUI} workflows, and troubleshooting runtime errors. \\

Broader AI Reflections & 11.2\% &
Covers posts about AI's social questions and ethical implications, industry news, and ChatGPT's limitations or jailbreaks. \\

Sharing, Feedback \& Community & 7.1\% &
Users share outputs, seek critique, or participate in contests. This theme is not about seeking general advice (Basic Setup \& Getting Started Help) or refining prompts (Prompt Practices \& Refinement), but about showcasing work and engaging in peer-driven evaluation and community validation. \\

Prompt Practices \& Refinement & 6.7\% &
Encompasses posts where users refine prompts or adjust prompting parameters to improve outputs, experiment with aesthetics, or fix issues like distorted hands.\\

\bottomrule
\end{tabular}
\end{table*}

\section{Results}

\subsection{Study 1: AI Discussion Themes in Online Creative Communities}

To answer RQ1, Study 1 maps the central themes of AI-related discussions in creative communities. The results reveal a strong emphasis on hands-on tool use, with reflection on capabilities, ethics, and community practices emerging only as secondary concerns.

\subsubsection{Interpreting Topic Modeling Results Through the Lens of AI Literacy}

Our post topic modeling results (see Table \ref{tab:post_topics} in Appendix \ref{appendix:topic_modeling}) revealed that, contrary to top-down accounts of AI literacy that emphasize abstract understanding, most discussions among creatives focused on the hands-on work of troubleshooting, prompting, and making AI tools function. Alongside this dominant concern with practice, we observed a smaller but steady current of reflection on capabilities, ethics, and community norms.  

Drawing on Long and Magerko (2020)'s definition of AI literacy \cite{durilong20}, we merged post topics into six themes (shown in Table \ref{tab:theme_differences}) that align with AI literacy dimensions. Rather than isolated categories, these themes form a recognizable progression: creators begin with setup and experimentation, confront obstacles, step back into broader reflection, and ultimately consolidate learning through community exchange. Together, these patterns highlight AI literacy in creative communities as an emergent, socially situated practice that develops through doing and sharing, not as abstract knowledge acquired in advance.

\textbf{Getting in the door: setup and first steps.} The \textit{Basic Setup \& Getting Started Help (23.8\%)} reflects broad, entry-level requests for guidance on applying AI tools. These posts capture the exploratory stage of AI literacy: tool recommendations, subscriptions, and initial setup. For many creators, literacy begins with overcoming basic barriers to adoption, often framed as ``help'' like ``can someone help...''.

\textbf{Experimenting and tinkering: making it work.} After gaining access to tools, creators turned to interactive refinement---adjusting prompts, customizing workflows and tuning models to meet creative goals. Two themes captured this activity: \textit{Prompting Practices \& Refinement theme (6.7\%) and \textit{Model Training \& Workflow Customization theme (13.3\%)}}. In the first, creators modified prompts and parameters to achieve desired aesthetics (e.g., anime, watercolor) or fix flaws such as distorted hands. The second one reflected more advanced engagement, including fine-tuning models (e.g., LoRA, DreamBooth), configuring ComfyUI node graphs, and troubleshooting errors. These practices reflect both the dialogic and practice-based dimensions of AI literacy: creators iteratively probe, interpret and refine the system's outputs while using AI as a tool to achieve their creative tasks.

\textbf{Confronting obstacles: frustrations and failures.} While experimenting, creators frequently encountered breakdowns. The largest theme, \textit{Tool-Related Complaints (26.2\%)}, captured widespread frustrations with unreliable outputs, distorted images, declining model quality, and restrictive platform policies such as \textit{Adobe}'s terms of service. These posts demonstrate how critical evaluation often arises through failure: literacy develops not just from learning what works, but from recognizing and diagnosing what does not.

\textbf{Stepping back: broader reflections.} A smaller but notable portion of discourse, \textit{Broader AI Reflections (11.2\%)}, moved beyond tools to consider systemic and ethical questions. Creators discussed industry policies (e.g., \textit{OpenAI} pricing), raised issues of bias and labor displacement, critiqued limitations (e.g., incoherent text), and described safeguard restrictions or jailbreaks (e.g., DAN, which stands for ``Do Anything Now''). These reflections often emerged alongside external developments---such as model launches, pricing changes, or from sustained concerns about the broader implications of AI. 

\textbf{Learning together: community validation and peer support.} Finally, the \textit{Sharing, Feedback \& Community theme (7.1\%)} highlights how AI literacy is collectively sustained. Creators shared outputs, entered contests, promoted projects, and sought critique---for example, requesting feedback on AI-generated portraits or showcasing work to inspire peers, treating community participation as both a validation mechanism and a peer-learning infrastructure. Here, literacy is not only about individual competence but about contributing to shared knowledge and norms. 

\subsubsection{Qualitative Conversations Analysis Results}

While topic modeling provided a bird's-eye view of the themes of AI discussions, our follow-up qualitative analysis of 900 sampled conversations offered a closer look at how these themes unfold in practice. The qualitative analysis not only corroborates the dominant clusters identified in the topic modeling (e.g., tool complaints, setup help) but also sharpens them into more specific forms of AI literacy, such as procedure-based help-seeking or capacity testing. From this, we identify eight themes, and four of them are literacy-related: tool literacy was by far the most prominent, followed by capacity awareness, ethics and responsible use, and community engagement (see Table \ref{tab:literacy_themes}).  
Together, these themes reveal how AI literacy is co-constructed in everyday exchanges, with tool-focused problem-solving forming the foundation for reflection and collective learning. We examine each literacy-related theme in detail below, drawing on representative excerpts from the dataset.

\begin{table*}[tb]
\centering
\caption{Overview of the four literacy-related themes identified through qualitative analysis. Percentages are calculated over all 810 AI-related conversations. Note: themes that are not directly related to AI literacy (AI Output Sharing, AI Industry Dynamics, and Promotion) are excluded, so the percentages do not sum to 100\%.}
\label{tab:literacy_themes}
\small
\setlength\tabcolsep{6pt}
\renewcommand{\arraystretch}{1.3}
\begin{tabular}{@{} p{4cm} p{1.5cm} p{8.5cm} @{}} 
\toprule
\textbf{Theme} & \textbf{Percentage} & \textbf{Description} \\
\midrule

Tool Literacy & 46.0\% &
Practical, hands-on engagement with GenAI tools, including setup, prompting, troubleshooting, workflow configuration, hardware issues, resource requests, and prompt tuning. Reflects creators’ dominant focus on “making the tools work.” \\

Capacity Awareness & 15.4\% &
How creators reason about AI capabilities and limitations through capacity testing, sharing failure cases, discussing internal mechanisms, and probing model behavior. Captures how users form mental models of what AI can and cannot do. \\

Ethics \& Responsible Use & 11.5\% &
Concerns related to the ethical and societal implications of AI, including labor impacts, fairness and bias, copyright, misuse, safety guardrails, data privacy, and AI lab policies. \\

Community Engagement & 9.9\% &
Collective learning practices such as resource sharing, workflow documentation, code snippets, tutorials, and peer feedback. \\

\bottomrule
\end{tabular}
\end{table*}

\textbf{Theme 1: Tool Literacy} We found that 46.0\% (372/810) of sampled conversations were related to creators developing practical competence in setting up, configuring, prompting and troubleshooting AI tools to achieve creative outcomes. This theme corresponds to the topic modeling themes of ``\textit{Tool Related Complaints}'',  ``\textit{Basic Setup \& Getting Started Help}'', ``\textit{Model Training \& Workflow Cus-
tomization}'' and ``\textit{Prompt Practices \& Refinement}'', which all reflect practical tool use. It illustrates how these high-level themes are enacted through concrete user practices. 

A majority of these conversations (284/372, 76.3\%) involved help-seeking behavior, ranging from basic setup procedures to more complex troubleshooting challenges. The remaining posts covered hardware configuration (17/372, 4.6\%), access and authentication issues (16/372, 4.3\%), tool comparisons (35/372, 9.4\%), and prompt sharing (20/372, 5.4\%). The dominance of help-seeking conversations prompted us to examine this category more closely. Drawing on findings from prior research \cite{someone, onlinehelp}, we developed additional subcategories to capture the specific types of assistance that creators were seeking: procedural help, determining possibilities, interpretive questions, troubleshooting, seeking resource recommendations, prompt feedback, and descriptive questions. Table~\ref {tab:help_seeking_types} presents each type of help-seeking behavior along with its percentage and corresponding examples. 

Of the help-seeking posts, over a quarter (84/284, 29.6\%) were procedure-based questions. Creators frequently asked step-by-step questions about accomplishing specific desired creative tasks, configuring tools or workflow inquiries. For instance, \textit{``I'm trying to run \textit{Stable Diffusion} with ControlNet on Replicate. I've set up Stable Diffusion and each ControlNet variant separately, but I can't figure out how to combine them into one pipeline. How can I do this?''}. In many cases, the author would describe their creative goals and ask for guidance on how to achieve them: \textit{``I uploaded a photo link to generate images, but the face changes too much in the results. What prompts or keywords can help keep the face consistent and realistic?''}. This category also included questions from newcomers asking how to get started with the tools.

\begin{table*}[tb]
  \centering
  \caption{Types of help-seeking conversations observed in the dataset, with percentages and representative examples.}
  \label{tab:help_seeking_types}
  \small
  \begin{tabular}{@{} p{4cm} p{1.5cm} p{8.5cm} @{}} 
    \toprule
    \textbf{Help-seeking Type} & \textbf{Percentage} & \textbf{Example} \\
    \midrule
    Procedure help (including newcomers asking for the first step to start)
      & 29.6\%
      & ``How do I do this with [the tool]? What do I do first?'' \\[0.5em]
    Determine possibilities
      & 22.5\%
      & ``Can I do this with the [tool]?'' \\[0.5em]
    Interpretive questions
      & 13.4\%
      & ``Why does this happen? What did I do wrong?'' \\[0.5em]
    Troubleshooting
      & 12.3\%
      & ``How do I fix it?'' \\[0.5em]
    Resource recommendation request (including seeking tips sharing and tool recommendations)
      & 9.9\%
      & ``Is there a tutorial that anyone can share?'' \\[0.5em]
    Prompt feedback
      & 9.5\%
      & ``I just can't seem to find a good prompt—even detailed prompts don't work with Stable Diffusion. Here is a 'before' image of what I want, and here is the 'after' image it gives me. Please help me.'' \\[0.5em]
    Descriptive questions
      & 2.8\%
      & ``What is this? What is the difference between...?'' \\
    \bottomrule
  \end{tabular}
\end{table*}

The second most frequent type was ``determining possibilities'' (64/284, 22.5\%)---asking whether specific creative tasks could be accomplished with available tools. For example, one post asked, \textit{``Most face enhancement tools work well when the subject is looking straight ahead, but they struggle when the face is turned or significantly angled. Are there any tools capable of handling faces that are looking away?''}

Interpretive questions (38/284, 13.4\%) and troubleshooting (35/284, 12.3\%) were nearly equal in frequency. Interpretive questions arose when users were confused about what caused a problematic result or sought explanations of why something behaves the way it does, such as \textit{``I trained Stable Diffusion on both me and my friend using fast-DreamBooth, but when I try to generate my face, the AI creates a mix of both our faces instead of just mine. What happened?''}. Troubleshooting occurs when users encounter unexpected behavior such as runtime errors with an AI tool and seek help resolving these issues. In many cases, these posts include code snippets, with authors hoping that others can identify and help fix the problems they're experiencing. Relatedly, resource recommendation requests (28/284, 9.9\%) occurred when users sought resources such as tutorials, tools, models, or tips. Prompt feedback (27/284, 9.5\%) involved users seeking feedback to improve their prompts or expressing uncertainty about what prompts to use to achieve their desired results, such as \textit{``I'm trying to make a firbolg character (a character in the Dungeons \& Dragons game) art in Midjourney, but have no idea how to word it. Any advise?''} The least frequent type was descriptive questions (8/284, 2.8\%), which involved posts seeking basic information or comparisons about AI tools.


\textbf{Theme 2: Capacity Awareness} The second most frequent theme was capacity awareness, referring to creators' reflections on model behavior, including both limitations and capabilities. About 15.4\% of the sample (125/810) fell into this category, encompassing four subtypes: limitation awareness (46/125, 36.8\%), capacity testing (26/125, 20.8\%), internal mechanism discussions (30/125, 24.0\%), and strength recognition (23/125, 18.4\%). 

Many posts (36.8\%) in this theme were from users documenting and sharing model limitations they had discovered. For example, one post solicited \textit{``simple prompts that ChatGPT-4 fails on''} for a blog article and offered one example: \textit{``How many words are in your response to this prompt?''}. Capacity testing appeared in 26/125 posts, with most being curiosity-driven explorations (19/26) where users probed models to observe outputs, while others involved task-specific assessments (7/26) evaluating whether models could complete particular tasks. Conversations in both subcategories commonly included example prompts, partial transcripts, or screenshots that showcased interesting outputs or unexpected failures. Curiosity-driven probes tended to adopt a playful tone; for instance, one post asked, \textit{``If I ask ChatGPT what it wants to be named, what will it say? Has anyone else tried this?''} By contrast, task-specific assessments were more goal-oriented, such as one post challenged ChatGPT to \textit{``write some music''} in guitar-TAB notation.

Internal mechanism exploration (30/125, 24.0\%) occurred when creators described or asked questions about how AI models work. For example, one post explored training data considerations: \textit{``I've searched a bit, but there doesn't seem to be much info out there. EXIF data from cameras feels like it could be a goldmine for training, but most public models seem trained on scraped images that likely don't include it. Has anyone tried adding this onto an existing model with a LoRA or something similar?''}. Users often sought to understand why models produced certain outputs or how training data influenced model capabilities. One post asked why DALLE always warps the text into a blend of gibberish and vaguely Scandinavian characters when generating a logo with text. Comments explained that, because AI is trained on images not text so it lacks understanding of how to integrate these two. 

Strength recognition (23/125, 18.4\%) captured posts where creators reflected on successful applications of AI tools, often documenting notable use cases or results. For instance, one post shared a practical success story: \textit{``I was saving a cover image for my digital recipe book, and for some reason it had a guy's face floating in the corner. I don't have Photoshop, so I asked ChatGPT to test it on 10/10 success.''} Unlike posts focused on limitations or failures, these emphasized the capabilities of the models and offered practical insights for others looking to take advantage of similar strengths.

\textbf{Theme 3: Ethics and Responsible Use} Ethics and responsible use emerged as a major theme, with 93 of 810 posts (11.5\%) addressing the ethical, legal, and safety implications of AI. Whereas topic modeling grouped these issues more broadly under the ``Broader AI Reflections'' cluster, our qualitative analysis disentangles specific concerns---such as copyright, bias, safety guardrails, and data privacy---revealing the concrete ways creators articulate ethics in practice.

The largest subcategory, impact of AI (36/93, 38.7\%), centered on effects on employment, creative work, and education. For instance, one post linked the Hollywood writers' strike to speculation about studios pairing writers with ChatGPT; another post summarized a news report on occupations at high risk of AI replacement; and another one asked how to counter claims that AI \textit{``steals''} artists' jobs by framing Stable Diffusion as a creative tool. 

Misuse and safety accounted for 17 of 93 ethics-related posts (18.3\%). This subcategory included critiques of safety measures for degrading output quality and attempts to bypass them to generate restricted content. For example, one post described how boilerplate \textit{``as an AI...''} responses consumed up to two-thirds of creative writing outputs, raising concerns that guardrails hinder tasks like novel writing or game design. 

Other ethical topics included bias (11/93, 11.8\%), where users raised concerns about fairness and representation in outputs. Copyright (8/93, 8.6\%) focused on ownership of AI-generated content and related policies; for example, one post sought cases of artists losing credit, rights, or payment due to AI appropriation. Comparisons with humans (8/93, 8.6\%) examined AI capabilities relative to human performance. Data privacy (7/93, 7.5\%) and AI lab policy (6/93, 6.5\%) rounded out the remaining ethical concerns.


\textbf{Theme 4: Community Engagement} Community Engagement emerged as another theme, appearing in 80 of the 810 conversations (9.9\%) where users actively contributed to collective knowledge sharing. Although less visible in the topic modeling stage, our qualitative analysis surfaces Community Engagement as a distinct literacy-building practice, highlighting how resource sharing, workflow documentation, and peer feedback sustain collective learning.

This theme comprised three subcategories that reflected different forms of participation and support. Nearly half of these conversations (38/80, 47.5\%) involved resource sharing. In these posts, users provided code snippets, tutorials, tools, models, blog posts, and other materials they had developed or discovered. For example, one post shared: \textit{``I fine-tuned a Stable Diffusion model with ControlNet to generate logos from text. Let me know if you find it useful. Hugging Face space link: [link details omitted].''} 

Workflow sharing (23/80, 28.8\%) involved users posting detailed descriptions of their processes and settings for using AI tools effectively in creative tasks. These posts ranged from outlining full pipelines to listing hyperparameter choices, enabling others to adopt, adapt, and build on their approaches. We distinguish workflow sharing from AI output sharing by the level of process detail provided.

Peer feedback accounted for the remaining conversations (19/80, 23.8\%), where community members posted works-in-progress or finished outputs and sought constructive critique from others.

Collectively, these modes of participation demonstrate how literacy is co-constructed through shared practice rather than developed in isolation.

\subsection{Study 2: Tracking How Themes Shift}

Our first study identified and refined conversation themes within creative communities, but offered only a static view. In Study 2, we adopt an exploratory approach to examine how these themes shift over time and alongside major AI events, such as tool releases and policy updates. 

For this analysis, we exclude ``AI Output Sharing'', ``Promotion'', and ``AI Tech Dynamics Sharing'', as these themes fall outside the scope of our primary research questions. As described in the methodology, we report both proportion change (change in share of total conversations) and relative change (percent change in raw volume).

\textbf{Note}: There is extensive methodological literature on causal inference in time-series analysis (e.g., Granger causality and related approaches \cite{grangerpaper, casualinference}). However, given the large-scale and heterogeneous nature of public forum data, many factors could plausibly influence shifts in discussion themes. We therefore exercise caution in making strong causal claims. Instead, we interpret the observed patterns as offering insight into the dynamics of how discussion themes shift alongside AI-related events.

\begin{figure*}[tb]
  \centering
  \includegraphics[width=1.0\linewidth]{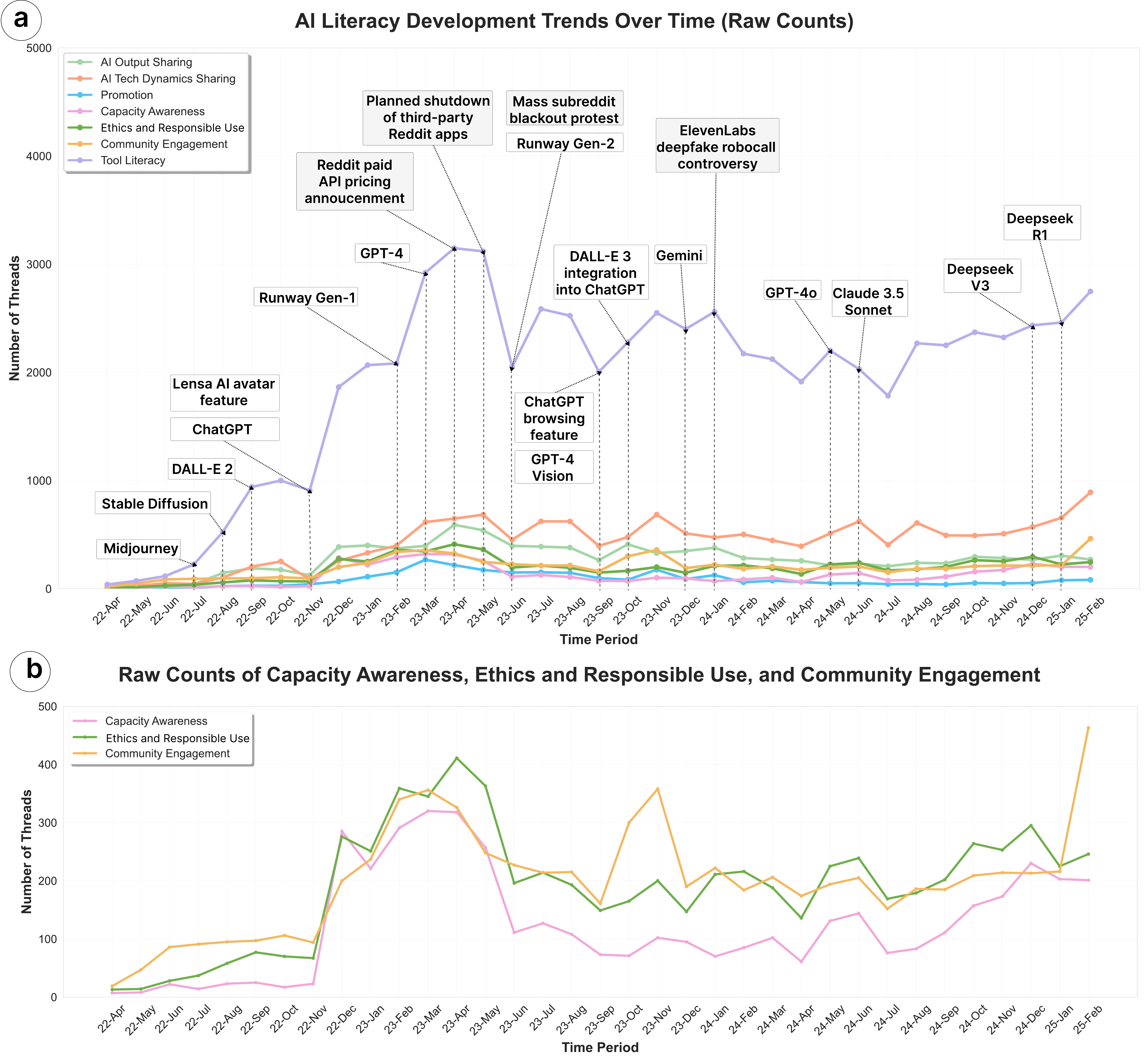}
  \caption{The top panel (a) shows the raw count of AI literacy conversations over time, with major AI tool releases, controversies, and platform events annotated. Key spikes align with high-impact moments such as the launch of ChatGPT. The bottom panel (b) isolates trends in Capacity Awareness, Ethics and Responsible Use, and Community Engagement, as these themes are often overshadowed by the dominant focus on Tool Literacy. \textit{Please note that subcharts (a) and (b) are using different y-axis scales.}}
  \label{fig:raw}
  \Description{The figure shows trends in AI literacy discussions over time using raw counts of Reddit threads. Panel a displays conversation volume across multiple themes, with Tool Literacy dominating and showing sharp increases aligned with major AI releases, platform changes, and controversies such as the launch of ChatGPT, GPT four, DALL E three integration, and Deepseek. Panel b focuses on Capacity Awareness, Ethics and Responsible Use, and Community Engagement using a different vertical scale, highlighting lower volume but recurring discussion patterns around major AI events. Together, the panels show that tool focused discussions drive overall volume, while reflective and social aspects of AI literacy appear intermittently.}
\end{figure*}

\begin{figure*}[htbp]
  \centering
  \includegraphics[width=1.0\linewidth]{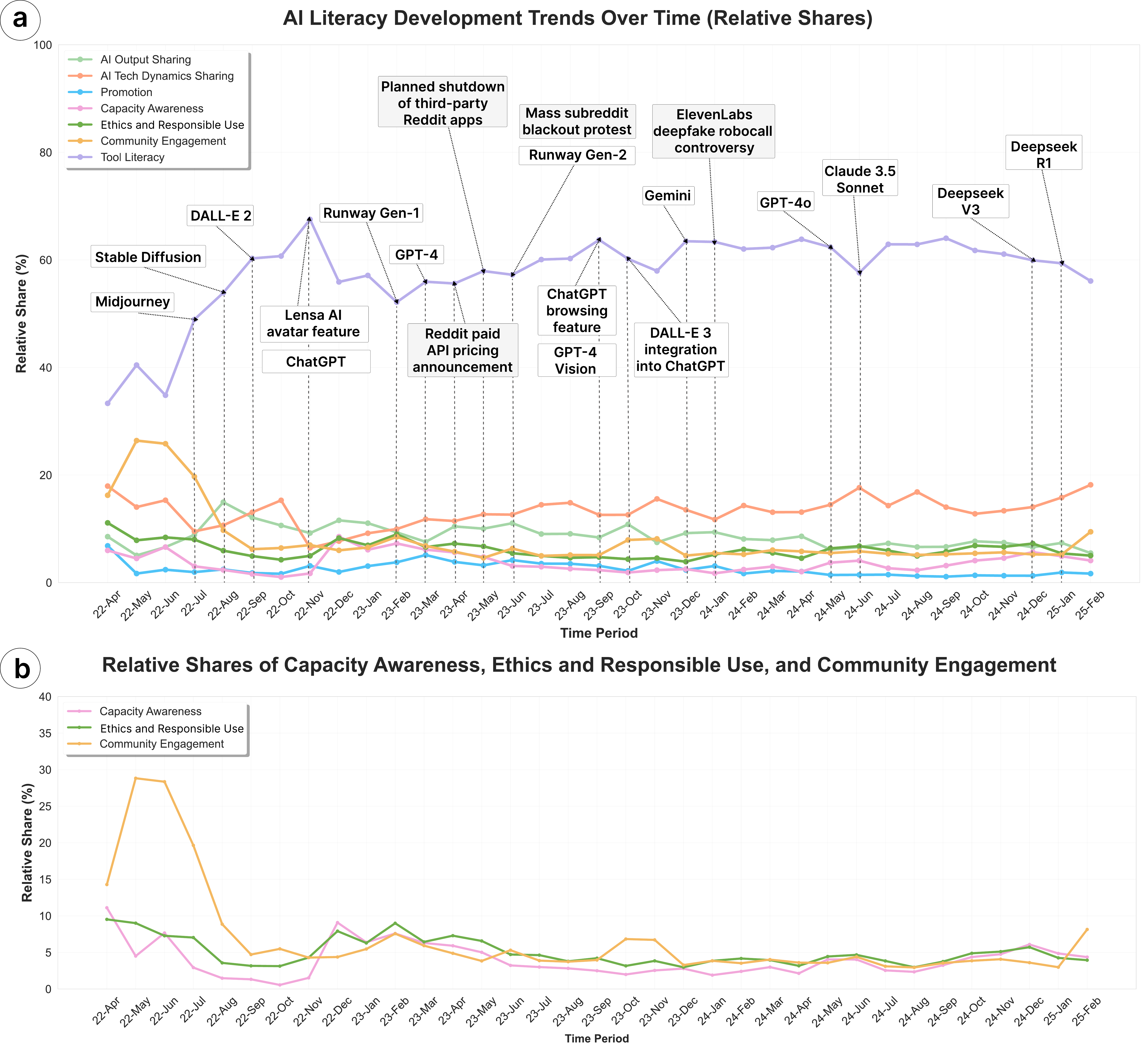}
  \caption{Trends in AI literacy discourse over time. Panel (a) shows the relative distribution of AI literacy conversations, annotated with major AI tool releases, controversies, and platform events. Tool Literacy remained the dominant theme throughout the observation period, accounting for approximately 55–60\% of discussions. A smaller share of conversations (around 4–7\%) is about capacity awareness and ethical considerations. Panel (b) isolates the relative shares of Capacity Awareness, Ethics and Responsible Use, and Community Engagement, highlighting their fluctuations over the same period. \textit{Note: Panels (a) and (b) use different y-axis scales.}}
  \label{fig:relative}
  \Description{The figure shows relative shares of AI literacy discussions over time. Panel a displays the proportion of conversations across themes, with Tool Literacy consistently accounting for the largest share, while other themes such as Capacity Awareness, Ethics and Responsible Use, Community Engagement, AI Output Sharing, and AI Industry Dynamics remain smaller but fluctuate around major AI releases, controversies, and platform events. Panel b focuses on Capacity Awareness, Ethics and Responsible Use, and Community Engagement using a different vertical scale, highlighting subtle shifts in their relative prominence over time. Together, the panels show that tool focused discussions dominate AI literacy discourse, while reflective and social themes occupy a smaller but persistent share.}
\end{figure*}

\textbf{Insight 1: Tool Literacy Dominates Conversations}. Tool Literacy consistently represented the largest category by both volume and proportion across all periods. A major expansion began in mid-2022 with the rise of text-to-image tools. Monthly conversations jumped from 39 in April to over 900 by November, driven by the popularity of Midjourney and Stable Diffusion. Midjourney's July release pushed the theme's share from 35\% in June to 54\% in August, while Stable Diffusion's August launch lifted it to a peak of 60\% in September (Figure~\ref{fig:relative}a). The launch of ChatGPT in November 2022 marked another turning point. While conversation volume nearly doubled from 1,000 in October to 1,864 in December, Tool Literacy's share declined slightly, signaling growing attention to other emerging themes.

From April to June 2023, Tool Literacy discussions dropped sharply (Figure~\ref{fig:raw}a) from 3,148 to 2,052 (-34.8\% relative decrease), coinciding with Reddit's API pricing changes that pressured third-party apps to shut down and triggered widespread protests \cite{protest1,protest2,protest3,apollo}. Protest messages---such as \textit{``your comments and posts are being sold by Reddit to Google to train AI...''}---amplified tensions and likely contributed to reduced engagement. Despite the volume drop, the theme's share rose by 1.6 percentage points (Figure~\ref{fig:relative}a), suggesting steeper declines in other themes.

To investigate whether the dominance of Tool Literacy is influenced by subreddit composition, we compared theme distributions across tool-specific communities (e.g., r/StableDiffusion, r/ChatGPT) and general creative communities (e.g., r/DigitalArt, r/ArtistLounge). Tool-specific subreddits exhibited a markedly higher concentration of Tool Literacy content, averaging 59.3\% of discussions (range: 23.8\%–72.6\%), and maintained this dominance over a three-year period. In contrast, general creative subreddits demonstrated a more balanced distribution of themes. While Tool Literacy still emerged as the most prevalent category overall, it accounted for a smaller average proportion of discussions (43.2\%; range: 23.8\%-59.7\%). This pattern suggests that while the prominence of tool-related discourse is partly shaped by community composition, practical tool use still constitutes a major component of how creators engage with AI across creative domains. The plots for each group, including raw counts and relative share, are provided in Appendix~\ref{appendix:plots}.

\textbf{Insight 2: Shifting Attention to Capacity Awareness and Model Capabilities} 
Capacity Awareness was initially a minor theme, but surged in November 2022 with the release of ChatGPT, which broadened the scope of AI discussions. Its share rose by 7.5 percentage points (Figure~\ref{fig:relative}b), with conversation volume increasing from 17 to 285 (+1,576.5\%) (Figure~\ref{fig:raw}b). Interest persisted in the subsequent months, with January 2023 reaching 221 conversations (+1,200.0\%) and February reaching 291 conversations (+1,611.8\%), indicating that attention to model capabilities extended beyond the initial launch surge.

This momentum carried into early 2023 before a sharp decline between April and June---coinciding with the Reddit API controversy (as previously noted \cite{protest1, protest2, protest3}): the raw number of conversations fell by over 65\% (Figure~\ref{fig:raw}b), and its share dropped by 2.5 percentage points (Figure~\ref{fig:relative}b). From July to December 2024, Capacity Awareness grew from 76 to 230 conversations. As of early 2025, activity remains at a high level, with 200+ conversations in January and February.

\textbf{Insight 3: Surges and Controversies in Ethics and Responsible Use}. Beginning with just 13 conversations in April 2022, Ethics and Responsible Use remained modest through November, reaching 67 posts. ChatGPT's November 2022 release nearly doubled the theme's share (from 4.3\% to 8.3\%) in Figure~\ref{fig:relative}b, with raw volume climbing from 70 in October to 276 in December (Figure~\ref{fig:raw}b). At the same time, \textit{Lensa AI'}s viral ``magic avatars'' feature introduced many users to paid AI-generated portraits, fueling debates about originality and artistic credit \cite {lensa2022}. 

A sharp decline followed during the Reddit API controversy, as conversations fell from 411 in April to 196 in June (-52.3\%). The theme rebounded in early 2024 following the \textit{ElevenLabs} deepfake controversy \cite{elevenlabs1, elevenlabs2}, rising from 147 in December 2023 to 216 in February 2024 (+46.9\%). Ethics and Responsible Use peaked at 295 posts in December 2024, dipped to 225 in January 2025, and rebounded to 246 in February, following the release of DeepSeek R1.

\textbf{Insight 4: Event-Driven Spikes in Community Engagement}
In April 2022, Community Engagement accounted for 19 posts. Although raw volume grew steadily to 106 by October, its share fell from 26.4\% to 6.4\%, as Tool Literacy surged during the rise of text-to-image tools (Figure~\ref{fig:raw}b and ~\ref{fig:relative}b). 

With ChatGPT's release in November 2022, Community Engagement doubled in volume, from 106 to 200 (+88.7\%), though the relative share still stayed low (Figure~\ref{fig:raw}b and ~\ref{fig:relative}b). A larger spike followed the October 2023 integration of DALL·E 3 into ChatGPT: conversations more than doubled from September to November (161->358), before dropping to 190 in December. Term frequency trends mirrored this surge: mentions of DALL·E 3 (or DALLE) in Community Engagement conversations climbed from 6 in September to 211 in October, then declined to 68 by December. ChatGPT mentions followed a similar arc, rising from 70 to 171 before declining to 62. In early 2025, Community Engagement volume rose again, adding 250 conversations (+117.4\%), coinciding with the release of DeepSeek R1 in January (Figure~\ref{fig:raw}b).

\textbf{Key Takeaways:} Tool Literacy accounted for over half of all conversations, reflecting a sustained focus on applying AI tools. Other themes---Capacity Awareness, Ethics, and Community Engagement gained attention during major events but remained secondary.


\section{Discussion}
By combining topic modeling with inductive analysis, this work contributes insights into (a) AI-related conversation themes emerging among creators on Reddit and, (b) how these themes evolve alongside major developments in AI. A key takeaway from our findings is that creators’ conversations primarily center on the practice-based dimensions of AI literacy. This pattern challenges expert-driven AI literacy frameworks (e.g., \cite{ng, kandlhofer2016, druga2019}) that position foundational knowledge as a prerequisite for being ``AI literate'' and also offers actionable insights for better supporting creators’ AI literacy development. Taken together, these findings suggest that AI literacy within online creative communities is neither uniform nor static---it emerges through hands-on tool use, community interaction, and shifts in external events such as high-profile model releases, tool updates and platform-level disruptions.


To move beyond description, we now turn to the broader implications of these findings for HCI: (1) conceptualizing AI literacy as an event-driven, social practice, (2) understanding the dominance of tool literacy and its implications, (3) examining the role of community-driven infrastructures in supporting learning, and (4) rethinking engagement with public AI discourse.

\subsection{AI Literacy as Event-Driven, Social Practice}

Our study shows that creators' AI literacy does not emerge all at once but evolves dynamically alongside major AI development events. We observed how discussions shifted from early questions about access and setup to more nuanced explorations of tool limitations, capacity testing, and ethical debates. Importantly, these shifts clustered around external triggers such as the release of Midjourney and Stable Diffusion, the launch of ChatGPT, or Reddit's API policy changes.

This responsiveness reframes AI literacy as a moving target. Rather than a static competency that individuals either have or lack, our findings highlight literacy as a socially situated and event-driven practice, unfolding in tandem with technological advances and policy shifts. In other words, literacy is not merely ``taught'' but co-constructed through ongoing exposure to new tools, their affordances, and the challenges they create.

For HCI, this suggests a conceptual contribution: literacy is best understood as embedded in sociotechnical infrastructures that are sensitive to disruption. This has implications for the design of learning resources, which should anticipate not only stable skills but also users' need to rapidly adapt when tools change.

\subsection{The Dominance of Tool Literacy in AI Conversations}

One of the most striking findings from our study is the dominance of tool literacy across conversations. Roughly 55–60\% of all AI-related posts focused on practical skills: installing and configuring tools, refining prompts, troubleshooting errors, and integrating models into creative workflows. This focus on ``making it work'' contrasts with prevailing definitions of AI literacy, which consistently frame it as a multi-dimensional construct encompassing not only technical use, but also conceptual understanding, critical evaluation, and ethical reflection \cite{ng,ngb,druga2019,druga2022,durilong20}. Compared to these frameworks, the community discourse we analyzed appears disproportionately oriented toward practical engagement, with more reflective and ethical dimensions present but less prominent.

Creators often expressed help-seeking behavior in highly specific, procedural ways---asking, for example, how to combine models, configure hardware, or fix distorted image outputs. 
These practices illustrate that for many, AI literacy begins not with abstract knowledge of algorithms but with applied, situated use cases.

This emphasis contrasts with prior frameworks that define AI literacy primarily in terms of understanding concepts, data processes, or ethical risks \cite{kandlhofer2016, wong2020, unicef, unesco, hermann2022}. While some researchers argue that failing to understand capacities and limitations risks unrealistic expectations \cite{yang2018, aiwildcard}, our findings suggest that starting with tool use is not a weakness but a practice-based entry point---such as building chatbots can provide opportunities for scaffolding AI literacy in practice \cite{shan_zhang_scaffolding_2025}. These examples suggest that educational and design interventions may be more effective if they scaffold outward from these practical engagements rather than assuming abstract knowledge as a prerequisite.

For HCI, the implication is clear: literacy-supporting systems should provide task-centered resources, contextual support, and debugging guidance that align with users' immediate goals, while gradually fostering awareness of limitations and ethical use. This reframing positions creators not as deficient but as engaged learners whose preferred mode of entry—tool literacy---can serve as a foundation for deeper understanding.

\subsection{Community-Driven Infrastructures of AI Literacy}

Public discourse often frames AI's impact on online communities in terms of decline, epitomized by the refrain that ``Stack Overflow is dying'' \cite{dead1,dead2}. Users are increasingly drawn to Gen AI tools (e.g., ChatGPT) because they provide immediate, well-articulated responses, even when accuracy remains uncertain \cite{delrio, samirakabir}. Additionally, some users feel ashamed or judged when posting on Stack Overflow, which may reinforce the shift toward Gen AI tools \cite{devguide, hackernews}.

Yet this decline narrative overlooks how AI tools have simultaneously generated new forms of community-based learning. Generative AI has enabled people without technical backgrounds to participate in creative domains such as image generation, but often pushes them into adjacent activities like coding and debugging, hardware configuration, and workflow scripting, particularly when running models locally. Many of these newcomers turn to Reddit, where they face fewer participation barriers. Posts often reflect socially embedded help-seeking practices---sharing code snippets, describing issues, and soliciting peer input. Prior work also suggests Reddit has avoided the same decline as Stack Overflow because it emphasizes social interaction and participation rather than purely knowledge exchange \cite{travis}.

These peer-learning interactions have evolved into sophisticated, bottom-up approaches to AI literacy. Communities deploy bot comments that prompt users to disclose inputs (e.g., \textit{``If your post is a ChatGPT screenshot, please reply with the conversation link or prompt. If it's a DALL·E image, reply with the prompt used.''}), which most often appear in posts about tool literacy or capacity awareness. Subreddits like \textit{r/StableDiffusion} also classify posts into categories such as Questions, Workflow Sharing and Tutorials to scaffold learning and facilitate critical evaluation.

Collectively, our findings suggest that AI literacy is fundamentally a social accomplishment rather than an individual competency. This aligns with Bandura's Social Learning Theory, which emphasizes that learning occurs through observation, modeling, and social interaction \cite{sociallearning}. The sophisticated practices we observed, from bot-enforced disclosure to collaboratively maintained taxonomies, communities are not witnessing the decline of collaborative knowledge sharing but its transformation into new forms shaped by emerging technologies. The social dimension of learning thus remains central, even as platforms and practices evolve.

\subsection{Rethinking Public Engagement with AI}

Finally, our findings highlight an important opportunity for the HCI, CS, and ML communities, policymakers, as well as companies developing AI systems, to engage more consistently and meaningfully with the public. As our findings show, what’s often framed as decline is better understood as a transformation---platforms like Reddit evolve into vibrant sites of peer learning about AI. 

Social media platforms such as Reddit and Twitter/X offer dynamic, unsolicited insights into how people interact with and make sense of AI in everyday life. These spaces allow researchers and practitioners to observe AI literacy as it unfolds in real-world contexts through hands-on tool use, peer learning, and ongoing discussion.

For HCI, this underscores the importance of studying emergent, public discourse rather than relying solely on surveys, expert frameworks, or formal curricula. Social platforms capture authentic, situated learning practices in real time, offering a lens into how diverse populations adapt to and make sense of new technologies.

Methodologically, our study contributes a roadmap for analyzing large-scale discourse to surface grassroots AI literacy. By combining topic modeling, LLM-assisted classification, and in-depth qualitative coding, we demonstrate how to trace literacy as it evolves over time alongside external events. This bottom-up approach complements top-down initiatives such as value surveys or curricular design by showing how literacy is produced ``from the ground up'' in everyday contexts. For example, OpenAI's Collective Alignment initiative surveyed over a thousand people to identify the values they believe AI systems should reflect \cite{alignment}. Likewise, an interview study with creative writers found that many expressed concern about their work being used as training data for LLMs without consent, revealing how creators articulate their values and expectations for AI systems \cite{gero2025creativewriters}. While such studies offer important normative input, they represent only one angle. Our findings underscore the value of pairing these top-down approaches with studies that examine how people make sense of AI through everyday discourse and problem-solving.

\section{Limitations}

While this study offers valuable insights into AI literacy discourse and the temporal evolution of AI-related conversations, several limitations should be acknowledged. First, this study focuses exclusively on Reddit, where subreddit-specific moderation policies shape the nature of discourse. Restrictions on AI-generated content often concentrate discussion within AI-focused communities \cite{travis}, potentially introducing topical bias. Moreover, factors such as subreddit selection bias (e.g., AI tool–specific subreddits that might naturally foreground tool discussions), Reddit’s platform affordances (e.g., greater visibility for help-seeking posts), and the tendency of more experienced creators to post less frequently may further influence the observed patterns.

Second, although LLMs enabled scalable classification, they may misinterpret nuanced language \cite{ambiguity, sarcasm, sentiment}, introducing classification noise. Third, our temporal analysis centers on major AI tool releases to preserve interpretive clarity, but this focus may overlook smaller or cumulative developments that also influence public discourse.

Future work could address these limitations by incorporating cross-platform data (e.g., Twitter/X, Discord), though this may be challenging to acquire, and expanding the scope beyond creative applications to include professional and educational AI use. Our findings suggest that AI literacy among creators is deeply rooted in tool use, emerging through hands-on experimentation and practical problem-solving. Building on this, future research might explore how onboarding experiences, contextual support, and workflow-specific resources can better align with users' goals and promote a deeper, situated understanding of AI tools.

\section{Conclusions}

In this paper, we examine how AI literacy emerges within online creative communities by analyzing large-scale public discourse over three years. Our findings show that AI literacy is neither static nor individually acquired, but develops as a bottom-up, event-driven, and socially situated practice shaped by workflow integration, help-seeking and troubleshooting, capability exploration, and ethical sense-making. By linking temporal patterns with the everyday practices that produce them, our work challenges expert-driven literacy frameworks and reveals the practical, community-based infrastructures through which creators learn to work with AI. 
These insights highlight opportunities to design systems and resources that better support the realities of situated, practice-based AI learning in everyday creative contexts.

\begin{acks}
We thank the Natural Sciences and Engineering Research Council of Canada (NSERC) for funding this research and Reddit for providing access to the dataset used in this study. We also thank Xueying Zhang and Aham Gupta for reading drafts of this paper and offering thoughtful feedback.

\end{acks}

\bibliographystyle{ACM-Reference-Format}
\bibliography{references}

@misc{lensa2022,
  author    = {Hatmaker, Taylor},
  title     = {Lensa AI, the app making ‘magic avatars,’ raises red flags for artists},
  year      = {2022},
  month     = dec,
  date      = {2022-12-05},
  url       = {https://techcrunch.com/2022/12/05/lensa-ai-app-store-magic-avatars-artists/},
  urldate   = {2025-09-10},
}

@misc{apollo,
  author    = {Perez, Sarah},
  title     = {Popular Reddit app Apollo may go out of business over Reddit’s new, unaffordable API pricing},
  year      = {2023},
  month     = may,
  date      = {2023-05-31},
  url       = {https://techcrunch.com/2023/05/31/popular-reddit-app-apollo-may-go-out-of-business-over-reddits-new-unaffordable-api-pricing/},
  urldate   = {2025-09-10},
}

@incollection{strauss1994grounded,
  author = {Strauss, Anselm and Corbin, Juliet},
  title = {Grounded Theory Methodology: An Overview},
  booktitle = {Handbook of Qualitative Research},
  editor = {Denzin, Norman K. and Lincoln, Yvonna S.},
  year = {1994},
  pages = {273--285},
  publisher = {SAGE Publications}
}

@misc{studyingreddit,
	title = {Studying {Reddit}: {A} {Systematic} {Overview} of {Disciplines}, {Approaches}, {Methods}, and {Ethics}},
	shorttitle = {Studying {Reddit}},
	url = {https://journals.sagepub.com/doi/epub/10.1177/20563051211019004},
	language = {en},
	urldate = {2025-09-08},
	year = {2021},
	doi = {10.1177/20563051211019004},
	journal = {Social Media + Society},
  	volume  = {6},
 	number  = {3},
	author = {Studer, Sebastian and Barbaro, Annette and Baur, Anna W. and Imhof, Christoph and Morger, Michael and Knecht, Michèle},

}

@inproceedings{chang2023promptartists,
  author    = {Chang, Minsuk and Druga, Stefania and Fiannaca, Alex and Vergani, Pedro and Kulkarni, Chinmay and Cai, Carrie and Terry, Michael},
  title     = {The Prompt Artists: Examining the Craft of Text‑to‑Image Model Users},
  booktitle = {Proceedings of the 2023 Creativity \& Cognition Conference},
  series    = {C\&C ’23},
  year      = {2023},
  publisher = {Association for Computing Machinery},
  address   = {New York, NY, USA},
  doi       = {10.1145/3591196.3593515},
  url       = {https://doi.org/10.1145/3591196.3593515}
}

@inproceedings{gero2025creativewriters,
  author    = {Gero, K. I. and West, Monica and Jakesch, Mollie and others},
  title     = {Creative Writers’ Attitudes on Writing as Training Data for Generative Models},
  booktitle = {Proceedings of the 2025 CHI Conference on Human Factors in Computing Systems},
  series    = {CHI ’25},
  year      = {2025},
  publisher = {Association for Computing Machinery},
  address   = {New York, NY, USA},
  doi       = {10.1145/3706598.3713287},
  url       = {https://doi.org/10.1145/3706598.3713287}
}

@misc{dead2,
	title = {Stack overflow is almost dead},
	url = {https://blog.pragmaticengineer.com/stack-overflow-is-almost-dead/},
	urldate = {2025-09-06},
	journal = {The Pragmatic Engineer},
	author = {Orosz, Gergely},
	month = may,
	year = {2025},
}

@misc{dead1,
	title = {Stack {Overflow} is dying: is it being replaced by {AI}?},
	shorttitle = {Stack {Overflow} is dying},
	url = {https://www.techzine.eu/news/devops/127669/stack-overflow-is-dying-is-it-being-replaced-by-ai/},
	language = {en},
	urldate = {2025-09-06},
	journal = {Techzine Global},
	author = {Klinken, Erik},
	month = jan,
	year = {2025},
}

@article{shan_zhang_scaffolding_2025,
	title = {Scaffolding {AI} {Literacy} {Through} {Student}-{AI} {Collaboration} in {Chatbot} {Development}},
	copyright = {Creative Commons Attribution 4.0 International},
	url = {https://zenodo.org/doi/10.5281/zenodo.15870131},
	doi = {10.5281/ZENODO.15870131},
	language = {en},
	urldate = {2025-09-06},
	author = {Shan Zhang and Anthony F. Botelho},
	editor = {Mills, Caitlin and Alexandron, Giora and Taibi, Davide and Lo Bosco, Giosuè and Paquette, Luc},
	month = jul,
	year = {2025},
}

@article{hermann2022,
	title = {Artificial intelligence and mass personalization of communication content—{An} ethical and literacy perspective},
	volume = {24},
	issn = {1461-4448},
	url = {https://doi.org/10.1177/14614448211022702},
	doi = {10.1177/14614448211022702},
	language = {EN},
	number = {5},
	urldate = {2025-09-01},
	journal = {New Media \& Society},
	author = {Hermann, Erik},
	month = may,
	year = {2022},
	pages = {1258--1277},
}

@article{laupichler2022,
	title = {Artificial intelligence literacy in higher and adult education: {A} scoping literature review},
	volume = {3},
	issn = {2666920X},
	shorttitle = {Artificial intelligence literacy in higher and adult education},
	url = {https://linkinghub.elsevier.com/retrieve/pii/S2666920X2200056X},
	doi = {10.1016/j.caeai.2022.100101},
	language = {en},
	urldate = {2025-09-01},
	journal = {Computers and Education: Artificial Intelligence},
	author = {Laupichler, Matthias Carl and Aster, Alexandra and Schirch, Jana and Raupach, Tobias},
	year = {2022},
	pages = {100101},
}

@article{kim2021,
	title = {Why and {What} to {Teach}: {AI} {Curriculum} for {Elementary} {School}},
	volume = {35},
	copyright = {Copyright (c) 2021 Association for the Advancement of Artificial Intelligence},
	issn = {2374-3468},
	shorttitle = {Why and {What} to {Teach}},
	url = {https://ojs.aaai.org/index.php/AAAI/article/view/17833},
	doi = {10.1609/aaai.v35i17.17833},
	language = {en},
	number = {17},
	urldate = {2025-09-01},
	journal = {Proceedings of the AAAI Conference on Artificial Intelligence},
	author = {Kim, Seonghun and Jang, Yeonju and Kim, Woojin and Choi, Seongyune and Jung, Heeseok and Kim, Soohwan and Kim, Hyeoncheol},
	month = may,
	year = {2021},
	pages = {15569--15576},
}

@article{wong_broadening_2020,
	title = {Broadening artificial intelligence education in {K}-12: where to start?},
	volume = {11},
	issn = {2153-2184, 2153-2192},
	shorttitle = {Broadening artificial intelligence education in {K}-12},
	url = {https://dl.acm.org/doi/10.1145/3381884},
	doi = {10.1145/3381884},
	language = {en},
	number = {1},
	urldate = {2025-09-01},
	journal = {ACM Inroads},
	author = {Wong, Gary K. W. and Ma, Xiaojuan and Dillenbourg, Pierre and Huan, John},
	month = feb,
	year = {2020},
	pages = {20--29},
}

@inproceedings{druga2022,
	address = {New York, NY, USA},
	series = {{CHI} '22},
	title = {Family as a {Third} {Space} for {AI} {Literacies}: {How} do children and parents learn about {AI} together?},
	isbn = {978-1-4503-9157-3},
	shorttitle = {Family as a {Third} {Space} for {AI} {Literacies}},
	url = {https://dl.acm.org/doi/10.1145/3491102.3502031},
	doi = {10.1145/3491102.3502031},
	urldate = {2025-08-31},
	booktitle = {Proceedings of the 2022 {CHI} {Conference} on {Human} {Factors} in {Computing} {Systems}},
	publisher = {Association for Computing Machinery},
	author = {Druga, Stefania and Christoph, Fee Lia and Ko, Amy J},
	month = apr,
	year = {2022},
	pages = {1--17},
}

@inproceedings{druga2019,
	address = {New York, NY, USA},
	series = {{FL2019}},
	title = {Inclusive {AI} literacy for kids around the world},
	isbn = {978-1-4503-6244-3},
	url = {https://dl.acm.org/doi/10.1145/3311890.3311904},
	doi = {10.1145/3311890.3311904},
	urldate = {2025-08-31},
	booktitle = {Proceedings of {FabLearn} 2019},
	publisher = {Association for Computing Machinery},
	author = {Druga, Stefania and Vu, Sarah T. and Likhith, Eesh and Qiu, Tammy},
	month = mar,
	year = {2019},
	pages = {104--111},
}

@inproceedings{kandlhofer2016,
	title = {Artificial intelligence and computer science in education: {From} kindergarten to university},
	shorttitle = {Artificial intelligence and computer science in education},
	url = {https://ieeexplore.ieee.org/document/7757570},
	doi = {10.1109/FIE.2016.7757570},
	urldate = {2025-09-01},
	booktitle = {2016 {IEEE} {Frontiers} in {Education} {Conference} ({FIE})},
	author = {Kandlhofer, Martin and Steinbauer, Gerald and Hirschmugl-Gaisch, Sabine and Huber, Petra},
	month = oct,
	year = {2016},
	pages = {1--9},
}

@misc{elevenlabs2,
	title = {{ElevenLabs} reportedly banned the account that deepfaked {Biden}'s voice with its {AI} tools},
	url = {https://ca.news.yahoo.com/elevenlabs-reportedly-banned-the-account-that-deepfaked-bidens-voice-with-its-ai-tools-083355975.html},
	author = {Moon, Mariella},
	language = {en-CA},
	urldate = {2025-08-29},
	journal = {Yahoo News},
	month = jan,
	year = {2024},
}

@article{elevenlabs1,
	title = {{AI} {Startup} {ElevenLabs} {Bans} {Account} {Blamed} for {Biden} {Audio} {Deepfake}},
	url = {https://www.bloomberg.com/news/articles/2024-01-26/ai-startup-elevenlabs-bans-account-blamed-for-biden-audio-deepfake},
	language = {en},
	urldate = {2025-08-29},
	journal = {Bloomberg.com},
	author = {Murphy, Margi and Metz, Rachel and Bergen, Mark},
	month = jan,
	year = {2024},
}

@misc{alignment,
	title = {Collective alignment: public input on our {Model} {Spec}},
	author ={OpenAI},
	shorttitle = {Collective alignment},
	url = {https://openai.com/index/collective-alignment-aug-2025-updates/},
	language = {en-US},
	urldate = {2025-08-29},
	month = aug,
	year = {2025},
}

@article{gencoglu_machine_2023,
	title = {Machine and expert judgments of student perceptions of teaching behavior in secondary education: {Added} value of topic modeling with big data},
	volume = {193},
	issn = {0360-1315},
	shorttitle = {Machine and expert judgments of student perceptions of teaching behavior in secondary education},
	url = {https://www.sciencedirect.com/science/article/pii/S0360131522002536},
	doi = {10.1016/j.compedu.2022.104682},
	urldate = {2025-08-24},
	journal = {Computers \& Education},
	author = {Gencoglu, Bilge and Helms-Lorenz, Michelle and Maulana, Ridwan and Jansen, Ellen P. W. A. and Gencoglu, Oguzhan},
	month = feb,
	year = {2023},
	pages = {104682},
}

@inproceedings{cheng_how_2022,
	address = {New Orleans LA USA},
	title = {How {Interest}-{Driven} {Content} {Creation} {Shapes} {Opportunities} for {Informal} {Learning} in {Scratch}: {A} {Case} {Study} on {Novices}’ {Use} of {Data} {Structures}},
	copyright = {https://www.acm.org/publications/policies/copyright\_policy\#Background},
	isbn = {978-1-4503-9157-3},
	shorttitle = {How {Interest}-{Driven} {Content} {Creation} {Shapes} {Opportunities} for {Informal} {Learning} in {Scratch}},
	url = {https://dl.acm.org/doi/10.1145/3491102.3502124},
	doi = {10.1145/3491102.3502124},
	language = {en},
	urldate = {2025-08-24},
	booktitle = {{CHI} {Conference} on {Human} {Factors} in {Computing} {Systems}},
	publisher = {ACM},
	author = {Cheng, Ruijia and Dasgupta, Sayamindu and Hill, Benjamin Mako},
	month = apr,
	year = {2022},
	pages = {1--16},
}

@misc{unicef,
	author = {{UNICEF Office of Global Insight and Policy}},
	title = {Policy {Guidance} on {AI} for {Children} (Version 2.0)},
	url = {https://www.unicef.org/globalinsight/media/2356/file/UNICEF-Global-Insight-policy-guidance-AI-children-2.0-2021.pdf},
	language = {en},
	urldate = {2025-08-22},
	year = {2021},

}

@article{wong2020,
	title = {Broadening artificial intelligence education in {K}-12: where to start?},
	volume = {11},
	issn = {2153-2184, 2153-2192},
	shorttitle = {Broadening artificial intelligence education in {K}-12},
	url = {https://dl.acm.org/doi/10.1145/3381884},
	doi = {10.1145/3381884},
	language = {en},
	number = {1},
	urldate = {2025-08-22},
	journal = {ACM Inroads},
	author = {Wong, Gary K. W. and Ma, Xiaojuan and Dillenbourg, Pierre and Huan, John},
	month = feb,
	year = {2020},
	pages = {20--29},
}

@article{aipainting,
	title = {Uncovering the {Evolution} of {Topics} about {AI} {Painting}: {Dynamic} {Topic} {Modeling} of 180k {Discourse} {Data} in an {Online} {Community}},
	copyright = {Creative Commons Attribution 4.0 International},
	shorttitle = {Uncovering the {Evolution} of {Topics} about {AI} {Painting}},
	url = {https://zenodo.org/doi/10.5281/zenodo.12729914},
	doi = {10.5281/ZENODO.12729914},
	language = {en},
	urldate = {2025-08-19},
	author = {Shiyao Wei and Ran Bi},
	editor = {Benjamin, Paaßen and Carrie, Demmans Epp},
	month = jul,
	year = {2024},
}

@inproceedings{oppenlaender2022,
	address = {Tampere Finland},
	title = {The {Creativity} of {Text}-to-{Image} {Generation}},
	isbn = {978-1-4503-9955-5},
	url = {https://dl.acm.org/doi/10.1145/3569219.3569352},
	doi = {10.1145/3569219.3569352},
	language = {en},
	urldate = {2025-08-19},
	booktitle = {Proceedings of the 25th {International} {Academic} {Mindtrek} {Conference}},
	publisher = {ACM},
	author = {Oppenlaender, Jonas},
	month = nov,
	year = {2022},
	pages = {192--202},
}

@book{king2017,
	title = {Designing with {Data}: {Improving} the {User} {Experience} with {A}/{B} {Testing}},
	isbn = {978-1-4493-3496-3},
	shorttitle = {Designing with {Data}},
	language = {en},
	publisher = {"O'Reilly Media, Inc."},
	author = {King, Rochelle and Churchill, Elizabeth F. and Tan, Caitlin},
	month = mar,
	year = {2017},
}

@inproceedings{yang2018,
	address = {Hong Kong China},
	title = {Investigating {How} {Experienced} {UX} {Designers} {Effectively} {Work} with {Machine} {Learning}},
	isbn = {978-1-4503-5198-0},
	url = {https://dl.acm.org/doi/10.1145/3196709.3196730},
	doi = {10.1145/3196709.3196730},
	language = {en},
	urldate = {2025-08-19},
	booktitle = {Proceedings of the 2018 {Designing} {Interactive} {Systems} {Conference}},
	publisher = {ACM},
	author = {Yang, Qian and Scuito, Alex and Zimmerman, John and Forlizzi, Jodi and Steinfeld, Aaron},
	month = jun,
	year = {2018},
	pages = {585--596},
}

@misc{hebron2016,
	title = {Machine {Learning} for {Designers}},
	url = {https://www.oreilly.com/library/view/machine-learning-for/9781491971444/},
	language = {en},
	urldate = {2025-08-19},
	journal = {O’Reilly Online Learning},
	author = {Hebron, Patrick}, 
	month = jun,
	year = {2016}, 
}

@article{huang_experiential_2023,
	title = {Experiential speculation in vision-based {AI} design education: {Designing} conventional and progressive {AI} futures},
	copyright = {Creative Commons Attribution 4.0 International},
	shorttitle = {Experiential speculation in vision-based {AI} design education},
	url = {http://www.ijdesign.org/index.php/IJDesign/article/view/4943},
	doi = {10.57698/V17I2.01},
	language = {en},
	urldate = {2025-08-19},
	author = {Huang, Janet Yi-Ching and Wensveen, Stephan and Funk, Mathias},
	year = {2023},
}

@inproceedings{aiwildcard,
	address = {Hamburg Germany},
	title = {{AI} {Is} {Not} a {Wildcard}: {Challenges} for {Integrating} {AI} into the {Design} {Curriculum}},
	isbn = {979-8-4007-0737-7},
	shorttitle = {{AI} {Is} {Not} a {Wildcard}},
	url = {https://dl.acm.org/doi/10.1145/3587399.3587410},
	doi = {10.1145/3587399.3587410},
	language = {en},
	urldate = {2025-08-19},
	booktitle = {Proceedings of the 5th {Annual} {Symposium} on {HCI} {Education}},
	publisher = {ACM},
	author = {Flechtner, Rahel and Stankowski, Aeneas},
	month = apr,
	year = {2023},
	pages = {72--77},
}

@article{ngb,
	title = {Artificial intelligence ({AI}) literacy education in secondary schools: a review},
	volume = {32},
	issn = {1049-4820},
	shorttitle = {Artificial intelligence ({AI}) literacy education in secondary schools},
	url = {https://doi.org/10.1080/10494820.2023.2255228},
	doi = {10.1080/10494820.2023.2255228},
	number = {10},
	urldate = {2025-08-19},
	journal = {Interactive Learning Environments},
	author = {Ng, Davy Tsz Kit and Su, Jiahong and Leung, Jac Ka Lok and Chu, Samuel Kai Wah},
	month = nov,
	year = {2024},
	pages = {6204--6224},
}

@article{ghallab,
	title = {Responsible {AI}: requirements and challenges},
	volume = {1},
	issn = {2523-398X},
	shorttitle = {Responsible {AI}},
	url = {https://doi.org/10.1186/s42467-019-0003-z},
	doi = {10.1186/s42467-019-0003-z},
	number = {1},
	urldate = {2025-08-18},
	journal = {AI Perspectives},
	author = {Ghallab, Malik},
	month = sep,
	year = {2019},
	pages = {3},
}

@article{Burgsteiner,
	title = {{IRobot}: {Teaching} the {Basics} of {Artificial} {Intelligence} in {High} {Schools}},
	volume = {30},
	copyright = {Copyright (c)},
	issn = {2374-3468},
	shorttitle = {{IRobot}},
	url = {https://ojs.aaai.org/index.php/AAAI/article/view/9864},
	doi = {10.1609/aaai.v30i1.9864},
	language = {en},
	number = {1},
	urldate = {2025-08-18},
	journal = {Proceedings of the AAAI Conference on Artificial Intelligence},
	author = {Burgsteiner, Harald and Kandlhofer, Martin and Steinbauer, Gerald},
	month = mar,
	year = {2016},
}

@article{biagini,
	title = {Towards an {AI}-{Literate} {Future}: {A} {Systematic} {Literature} {Review} {Exploring} {Education}, {Ethics}, and {Applications}},
	issn = {1560-4306},
	shorttitle = {Towards an {AI}-{Literate} {Future}},
	url = {https://doi.org/10.1007/s40593-025-00466-w},
	doi = {10.1007/s40593-025-00466-w},
	language = {en},
	urldate = {2025-08-18},
	journal = {International Journal of Artificial Intelligence in Education},
	author = {Biagini, Gabriele},
	month = mar,
	year = {2025},
}

@misc{CoRemix,
	title = {{CoRemix}: {Supporting} {Informal} {Learning} in {Scratch} {Community} {With} {Visual} {Graph} and {Generative} {AI}},
	shorttitle = {{CoRemix}},
	url = {http://arxiv.org/abs/2412.05559},
	doi = {10.48550/arXiv.2412.05559},
	urldate = {2025-08-12},
	publisher = {arXiv},
	author = {Chen, Yunnong and Shen, Yishu and Liu, Ruiyi and Yu, Xinyu and Sun, Lingyun and Chen, Liuqing},
	month = dec,
	year = {2024},
}

@inproceedings{Kim2016Mosaic,
	address = {Portland Oregon USA},
	title = {Mosaic: {Designing} {Online} {Creative} {Communities} for {Sharing} {Works}-in-{Progress}},
	isbn = {978-1-4503-4335-0},
	shorttitle = {Mosaic},
	url = {https://dl.acm.org/doi/10.1145/2998181.2998195},
	doi = {10.1145/2998181.2998195},
	language = {en},
	urldate = {2025-08-11},
	booktitle = {Proceedings of the 2017 {ACM} {Conference} on {Computer} {Supported} {Cooperative} {Work} and {Social} {Computing}},
	publisher = {ACM},
	author = {Kim, Joy and Agrawala, Maneesh and Bernstein, Michael S.},
	month = feb,
	year = {2017},
	pages = {246--258},
}

@article{Fiebrink2019,
	title = {Machine {Learning} {Education} for {Artists}, {Musicians}, and {Other} {Creative} {Practitioners}},
	volume = {19},
	issn = {1946-6226},
	url = {https://dl.acm.org/doi/10.1145/3294008},
	doi = {10.1145/3294008},
	language = {en},
	number = {4},
	urldate = {2025-08-11},
	journal = {ACM Transactions on Computing Education},
	author = {Fiebrink, Rebecca},
	month = dec,
	year = {2019},
	pages = {1--32},
}

@article{oppenlaender2024prompting,
	title = {Prompting {AI} {Art}: {An} {Investigation} into the {Creative} {Skill} of {Prompt} {Engineering}},
	year = {2024},
	volume = {0},
	month = nov,
	issn = {1044-7318},
	shorttitle = {Prompting {AI} {Art}},
	url = {https://doi.org/10.1080/10447318.2024.2431761},
	doi = {10.1080/10447318.2024.2431761},
	number = {0},
	urldate = {2025-08-11},
	journal = {International Journal of Human–Computer Interaction},
	author = {Oppenlaender, Jonas and Linder, Rhema and Silvennoinen, Johanna},
	pages = {1--23},
}

@article{sociallearning,
	title = {Social {Learning} {Theory}},
	language = {EN},
	urldate = {2025-08-11},
	author = {Bandura, Albert},
	month = sep,
	year = {1977},
	url ={https://www.asecib.ase.ro/mps/Bandura_SocialLearningTheory.pdf},
}

@misc{hackernews,
	author = {{Hacker News}},
	title = {Ask {HN}: {Why} {Is} {Stack} {Overflow} {Fading} {Away}? {\textbar} {Hacker} {News}},
	url = {https://news.ycombinator.com/item?id=41364798},
	urldate = {2025-08-10},
	year = {2024},
}

@misc{devguide,
	title = {Is {Stack} {Overflow} dying?” a dev’s guide to the decline, drama, and data},
	author = {Medium},
	shorttitle = {Is {Stack} {Overflow} dying?},
	url = {https://medium.com/@dev_tips/is-stack-overflow-dying-a-devs-guide-to-the-decline-drama-and-data-3361818fb703},
	language = {en},
	urldate = {2025-08-10},
	journal = {Medium},
	author = {{\textless}devtips/{\textgreater}},
	month = may,
	year = {2025},
}

@article{delrio,
	title = {Large language models reduce public knowledge sharing on online {Q}\&{A} platforms},
	volume = {3},
	issn = {2752-6542},
	url = {https://doi.org/10.1093/pnasnexus/pgae400},
	doi = {10.1093/pnasnexus/pgae400},
	number = {9},
	urldate = {2025-08-10},
	journal = {PNAS Nexus},
	author = {del Rio-Chanona, R Maria and Laurentsyeva, Nadzeya and Wachs, Johannes},
	month = sep,
	year = {2024},
	pages = {pgae400},
}

@inproceedings{samirakabir,
	title = {Is {Stack} {Overflow} {Obsolete}? {An} {Empirical} {Study} of the {Characteristics} of {ChatGPT} {Answers} to {Stack} {Overflow} {Questions}},
	shorttitle = {Is {Stack} {Overflow} {Obsolete}?},
	url = {http://arxiv.org/abs/2308.02312},
	doi = {10.1145/3613904.3642596},
	urldate = {2025-08-10},
	booktitle = {Proceedings of the {CHI} {Conference} on {Human} {Factors} in {Computing} {Systems}},
	author = {Kabir, Samia and Udo-Imeh, David N. and Kou, Bonan and Zhang, Tianyi},
	month = may,
	year = {2024},
	pages = {1--17},
}

@misc{protest3,
	title = {Reddit will charge hefty fees to the many third-party apps that access its data},
	url = {https://www.cnbc.com/2023/06/01/reddit-eyeing-ipo-charge-millions-in-fees-for-third-party-api-access.html},
	language = {en},
	urldate = {2025-08-09},
	journal = {CNBC},
	author = {Goswami, Rohan},
	month = jun,
	year = {2023},
}

@misc{protest2,
	title = {Reddit {Will} {Start} {Charging} {Big} {Companies} for {API} {Access}},
	url = {https://gizmodo.com/reddit-will-charge-big-companies-for-ai-api-access-1850350291},
	language = {en-US},
	urldate = {2025-08-09},
	journal = {Gizmodo},
	author = {Main, Nikki},
	month = apr,
	year = {2023},
}

@article{protest1,
	title = {Reddit is facing a major protest from its own moderators},
	url = {https://www.cbc.ca/news/business/reddit-blackout-1.6873756},
	language = {en-CA},
	urldate = {2025-08-09},
	journal = {CBC News},
	author = {Wong, Aloysius},
	month = jun,
	year = {2023},
}

@article{casualinference,
	title = {Causal inference for time series},
	volume = {4},
	copyright = {2023 Springer Nature Limited},
	issn = {2662-138X},
	url = {https://www.nature.com/articles/s43017-023-00431-y},
	doi = {10.1038/s43017-023-00431-y},
	language = {en},
	number = {7},
	urldate = {2025-08-09},
	journal = {Nature Reviews Earth \& Environment},
	author = {Runge, Jakob and Gerhardus, Andreas and Varando, Gherardo and Eyring, Veronika and Camps-Valls, Gustau},
	month = jul,
	year = {2023},
	pages = {487--505},
}

@article{grangerpaper,
	title = {Investigating {Causal} {Relations} by {Econometric} {Models} and {Cross}-spectral {Methods}},
	volume = {37},
	issn = {0012-9682},
	url = {https://www.jstor.org/stable/1912791},
	doi = {10.2307/1912791},
	number = {3},
	urldate = {2025-08-09},
	journal = {Econometrica},
	author = {Granger, C. W. J.},
	year = {1969},
	pages = {424--438},
}

@misc{sentiment,
	title = {Improving {In}-{Context} {Learning} with {Prediction} {Feedback} for {Sentiment} {Analysis}},
	url = {http://arxiv.org/abs/2406.02911},
	doi = {10.48550/arXiv.2406.02911},
	urldate = {2025-08-06},
	publisher = {arXiv},
	author = {Xu, Hongling and Wang, Qianlong and Zhang, Yice and Yang, Min and Zeng, Xi and Qin, Bing and Xu, Ruifeng},
	month = jun,
	year = {2024},
}

@misc{sarcasm,
	title = {{SarcasmBench}: {Towards} {Evaluating} {Large} {Language} {Models} on {Sarcasm} {Understanding}},
	shorttitle = {{SarcasmBench}},
	url = {http://arxiv.org/abs/2408.11319},
	doi = {10.48550/arXiv.2408.11319},
	urldate = {2025-08-06},
	publisher = {arXiv},
	author = {Zhang, Yazhou and Zou, Chunwang and Lian, Zheng and Tiwari, Prayag and Qin, Jing},
	month = aug,
	year = {2024},
}

@misc{ambiguity,
	title = {Do {LLMs} {Understand} {Ambiguity} in {Text}? {A} {Case} {Study} in {Open}-world {Question} {Answering}},
	shorttitle = {Do {LLMs} {Understand} {Ambiguity} in {Text}?},
	url = {http://arxiv.org/abs/2411.12395},
	doi = {10.48550/arXiv.2411.12395},
	urldate = {2025-08-06},
	publisher = {arXiv},
	author = {Keluskar, Aryan and Bhattacharjee, Amrita and Liu, Huan},
	month = nov,
	year = {2024},
}

@article{sachit,
	title = {The democratization dilemma: {When} everyone is an expert, who do we trust?},
	volume = {12},
	issn = {2662-9992},
	shorttitle = {The democratization dilemma},
	url = {https://www.nature.com/articles/s41599-025-04734-x},
	doi = {10.1057/s41599-025-04734-x},
	language = {en},
	number = {1},
	urldate = {2025-08-02},
	journal = {Humanities and Social Sciences Communications},
	author = {Mahajan, Sachit},
	month = mar,
	year = {2025},
	pages = {455},
}

@article{gagan,
	title = {Updates in {Human}-{AI} {Teams}: {Understanding} and {Addressing} the {Performance}/{Compatibility} {Tradeoff}},
	volume = {33},
	copyright = {https://www.aaai.org},
	issn = {2374-3468, 2159-5399},
	shorttitle = {Updates in {Human}-{AI} {Teams}},
	url = {https://ojs.aaai.org/index.php/AAAI/article/view/4087},
	doi = {10.1609/aaai.v33i01.33012429},
	language = {en},
	number = {01},
	urldate = {2025-07-20},
	journal = {Proceedings of the AAAI Conference on Artificial Intelligence},
	author = {Bansal, Gagan and Nushi, Besmira and Kamar, Ece and Weld, Daniel S. and Lasecki, Walter S. and Horvitz, Eric},
	month = jul,
	year = {2019},
	pages = {2429--2437},
}

@article{javier,
	title = {Incorporating {Data} {Literacy} into {Information} {Literacy} {Programs}: {Core} {Competencies} and {Contents}},
	volume = {63},
	copyright = {De Gruyter expressly reserves the right to use all content for commercial text and data mining within the meaning of Section 44b of the German Copyright Act.},
	issn = {1865-8423},
	shorttitle = {Incorporating {Data} {Literacy} into {Information} {Literacy} {Programs}},
	url = {https://www.degruyterbrill.com/document/doi/10.1515/libri-2013-0010/html},
	doi = {10.1515/libri-2013-0010},
	language = {en},
	number = {2},
	urldate = {2025-07-19},
	journal = {Libri},
	author = {Prado, Javier Calzada and Marzal, Miguel Ángel},
	month = jun,
	year = {2013},
	pages = {123--134},
}

@misc{medialiteracy,
	title = {Critical {Media} {Literacy}: {Crucial} {Policy} {Choices} for a {Twenty}-{First}-{Century} {Democracy}},
	shorttitle = {Critical {Media} {Literacy}},
	url = {https://journals.sagepub.com/doi/epdf/10.2304/pfie.2007.5.1.59},
	doi = {10.2304/pfie.2007.5.1.59},
	journal = {Policy Futures in Education},
	volume = {5},
	number = {1},
	pages = {59-69},
	language = {en},
	urldate = {2025-07-19},
	author = {Kellner, Douglas and Share, Jeff},
	year = {2007},
}

@article{static,
	title = {{AI} literacy for users – {A} comprehensive review and future research directions of learning methods, components, and effects},
	volume = {2},
	issn = {2949-8821},
	url = {https://www.sciencedirect.com/science/article/pii/S2949882124000227},
	doi = {10.1016/j.chbah.2024.100062},
	number = {1},
	urldate = {2025-07-17},
	journal = {Computers in Human Behavior: Artificial Humans},
	author = {Pinski, Marc and Benlian, Alexander},
	month = jan,
	year = {2024},
	pages = {100062},
}

@incollection{scratch,
	address = {Cham},
	title = {Supporting {Diverse} and {Creative} {Collaboration} in the {Scratch} {Online} {Community}},
	isbn = {978-3-319-13535-9 978-3-319-13536-6},
	url = {http://link.springer.com/10.1007/978-3-319-13536-6_12},
	language = {en},
	urldate = {2025-07-06},
	booktitle = {Mass {Collaboration} and {Education}},
	publisher = {Springer International Publishing},
	author = {Roque, Ricarose and Rusk, Natalie and Resnick, Mitchel},
	editor = {Cress, Ulrike and Moskaliuk, Johannes and Jeong, Heisawn},
	year = {2016},
	pages = {241--256},
}

@inproceedings{someone,
	address = {Dunedin, New Zealand},
	title = {“{I} {Would} {Just} {Ask} {Someone}”: {Learning} {Feature}-{Rich} {Design} {Software} in the {Modern} {Workplace}},
	copyright = {https://ieeexplore.ieee.org/Xplorehelp/downloads/license-information/IEEE.html},
	isbn = {978-1-7281-6901-9},
	shorttitle = {“{I} {Would} {Just} {Ask} {Someone}”},
	url = {https://ieeexplore.ieee.org/document/9127288/},
	doi = {10.1109/VL/HCC50065.2020.9127288},
	language = {en},
	urldate = {2025-06-20},
	booktitle = {2020 {IEEE} {Symposium} on {Visual} {Languages} and {Human}-{Centric} {Computing} ({VL}/{HCC})},
	publisher = {IEEE},
	author = {Kiani, Kimia and Chilana, Parmit K. and Bunt, Andrea and Grossman, Tovi and Fitzmaurice, George},
	month = aug,
	year = {2020},
	pages = {1--10},
}

@incollection{onlinehelp,
	address = {San Francisco, CA, USA},
	title = {Building user-centered on-line help},
	isbn = {978-1-55860-246-5},
	urldate = {2025-06-20},
	booktitle = {Human-computer interaction: toward the year 2000},
	publisher = {Morgan Kaufmann Publishers Inc.},
	author = {Sellen, Abigail and Nicol, Anne},
	month = jun,
	year = {1995},
	pages = {718--723},
}

@misc{federico,
	title = {Generative {AI} and the {Creative} {Industry}: {Finding} {Balance} {Between} {Apologists} and {Critics}},
	shorttitle = {Generative {AI} and the {Creative} {Industry}},
	url = {https://medium.com/@fdonelli/generative-ai-and-the-creative-industry-finding-balance-between-apologists-and-critics-686f449862fc},
	language = {en},
	urldate = {2025-06-18},
	journal = {Medium},
	author = {Donelli, Federico},
	month = sep,
	year = {2024},
}

@inproceedings{kamila2023,
	address = {Koli Finland},
	title = {{AI} {Competencies} for non-computer science students in undergraduate education: {Towards} a competency framework},
	isbn = {979-8-4007-1653-9},
	shorttitle = {{AI} {Competencies} for non-computer science students in undergraduate education},
	url = {https://dl.acm.org/doi/10.1145/3631802.3631829},
	doi = {10.1145/3631802.3631829},
	language = {en},
	urldate = {2025-06-10},
	booktitle = {Proceedings of the 23rd {Koli} {Calling} {International} {Conference} on {Computing} {Education} {Research}},
	publisher = {ACM},
	author = {Tenório, Kamilla and Romeike, Ralf},
	month = nov,
	year = {2023},
	pages = {1--12},
}

@inproceedings{xie2025,
	address = {Yokohama Japan},
	title = {Exploring {What} {People} {Need} to {Know} to be {AI} {Literate}: {Tailoring} for a {Diversity} of {AI} {Roles} and {Responsibilities}},
	isbn = {979-8-4007-1394-1},
	shorttitle = {Exploring {What} {People} {Need} to {Know} to be {AI} {Literate}},
	url = {https://dl.acm.org/doi/10.1145/3706598.3713841},
	doi = {10.1145/3706598.3713841},
	language = {en},
	urldate = {2025-06-10},
	booktitle = {Proceedings of the 2025 {CHI} {Conference} on {Human} {Factors} in {Computing} {Systems}},
	publisher = {ACM},
	author = {Xie, Shixian and Zimmerman, John and Eslami, Motahhare},
	month = apr,
	year = {2025},
	pages = {1--16},
}

@misc{unesco,
	author = {UNESCO},
	title = {{AI} competency framework for students - {UNESCO} {Digital} {Library}},
	url = {https://unesdoc.unesco.org/ark:/48223/pf0000391105},
	urldate = {2025-05-18},
	year = {2024},
}

@misc{report,
	title = {How the {U}.{S}. {Public} and {AI} {Experts} {View} {Artificial} {Intelligence}},
	url = {https://www.pewresearch.org/internet/2025/04/03/how-the-us-public-and-ai-experts-view-artificial-intelligence/},
	language = {en-US},
	urldate = {2025-05-14},
	journal = {Pew Research Center},
	author = {Pasquini, Brian Kennedy, Jeffrey Gottfried, Monica Anderson {and} Giancarlo, Colleen McClain},
	month = apr,
	year = {2025},
}

@article{kappa,
	title = {The measurement of observer agreement for categorical data},
	volume = {33},
	issn = {0006-341X},
	language = {eng},
	number = {1},
	journal = {Biometrics},
	author = {Landis, J. R. and Koch, G. G.},
	month = mar,
	year = {1977},
	pages = {159--174},
}

@inproceedings{npmi,
	address = {Potsdam, Germany},
	title = {Evaluating {Topic} {Coherence} {Using} {Distributional} {Semantics}},
	url = {https://aclanthology.org/W13-0102/},
	urldate = {2025-04-26},
	booktitle = {Proceedings of the 10th {International} {Conference} on {Computational} {Semantics} ({IWCS} 2013) – {Long} {Papers}},
	publisher = {Association for Computational Linguistics},
	author = {Aletras, Nikolaos and Stevenson, Mark},
	editor = {Koller, Alexander and Erk, Katrin},
	month = mar,
	year = {2013},
	pages = {13--22},
}

@article{blei,
	title = {Latent dirichlet allocation},
	volume = {3},
	issn = {1532-4435},
	number = {null},
	journal = {J. Mach. Learn. Res.},
	author = {Blei, David M. and Ng, Andrew Y. and Jordan, Michael I.},
	month = mar,
	year = {2003},
	pages = {993--1022},
}

@misc{gensim,
	title = {Gensim: topic modelling for humans},
	shorttitle = {Gensim},
	author = {Rehurek, Radim},
	url = {https://radimrehurek.com/gensim/parsing/preprocessing.html#gensim.parsing.preprocessing.STOPWORDS},
	language = {en},
	urldate = {2025-04-24},
	note = {Accessed: 2025-09-09},
	year = {2025},
}

@misc{reddit4,
	type = {Reddit {Post}},
	title = {The {Reddit} for {Researchers} {Beta} {Program} is {Growing}!},
	url = {https://www.reddit.com/r/TheoryOfReddit/comments/1g62otb/the_reddit_for_researchers_beta_program_is_growing/},
	urldate = {2025-04-12},
	journal = {r/TheoryOfReddit},
	author = {PeerRevue},
	month = oct,
	year = {2024},
}

@article{chatgpt,
	title = {The public attitude towards {ChatGPT} on reddit: {A} study based on unsupervised learning from sentiment analysis and topic modeling},
	volume = {19},
	issn = {1932-6203},
	shorttitle = {The public attitude towards {ChatGPT} on reddit},
	url = {https://www.ncbi.nlm.nih.gov/pmc/articles/PMC11093324/},
	doi = {10.1371/journal.pone.0302502},
	number = {5},
	urldate = {2025-04-11},
	journal = {PLOS ONE},
	author = {Xu, Zhaoxiang and Fang, Qingguo and Huang, Yanbo and Xie, Mingjian},
	month = may,
	year = {2024},
	pages = {e0302502},
}

@inproceedings{cvmeasure,
	address = {Shanghai China},
	title = {Exploring the {Space} of {Topic} {Coherence} {Measures}},
	isbn = {978-1-4503-3317-7},
	url = {https://dl.acm.org/doi/10.1145/2684822.2685324},
	doi = {10.1145/2684822.2685324},
	language = {en},
	urldate = {2025-04-03},
	booktitle = {Proceedings of the {Eighth} {ACM} {International} {Conference} on {Web} {Search} and {Data} {Mining}},
	publisher = {ACM},
	author = {Röder, Michael and Both, Andreas and Hinneburg, Alexander},
	month = feb,
	year = {2015},
	pages = {399--408},
}

@inproceedings{umass,
	address = {Edinburgh, Scotland, UK.},
	title = {Optimizing {Semantic} {Coherence} in {Topic} {Models}},
	url = {https://aclanthology.org/D11-1024/},
	urldate = {2025-04-03},
	booktitle = {Proceedings of the 2011 {Conference} on {Empirical} {Methods} in {Natural} {Language} {Processing}},
	publisher = {Association for Computational Linguistics},
	author = {Mimno, David and Wallach, Hanna and Talley, Edmund and Leenders, Miriam and McCallum, Andrew},
	editor = {Barzilay, Regina and Johnson, Mark},
	month = jul,
	year = {2011},
	pages = {262--272},
}

@misc{stopword2,
	title = {Topic {Modeling} for the {People}},
	url = {https://maria-antoniak.github.io//2022/07/27/topic-modeling-for-the-people.html},
	author = {Antoniak, Maria},
	language = {en},
	urldate = {2025-04-02},
	month = mar,
	year = {2023},
}

@inproceedings{stopword1,
	address = {Valencia, Spain},
	title = {Pulling {Out} the {Stops}: {Rethinking} {Stopword} {Removal} for {Topic} {Models}},
	shorttitle = {Pulling {Out} the {Stops}},
	url = {https://aclanthology.org/E17-2069/},
	urldate = {2025-04-02},
	booktitle = {Proceedings of the 15th {Conference} of the {European} {Chapter} of the {Association} for {Computational} {Linguistics}: {Volume} 2, {Short} {Papers}},
	publisher = {Association for Computational Linguistics},
	author = {Schofield, Alexandra and Magnusson, Måns and Mimno, David},
	editor = {Lapata, Mirella and Blunsom, Phil and Koller, Alexander},
	month = apr,
	year = {2017},
	pages = {432--436},
}

@book{towardsailiteracy,
	address = {Calgary},
	title = {Towards {AI} {Literacy}: 101+ {Creative} and {Critical} {Practices}, {Perspectives} and {Purposes}},
	shorttitle = {Towards {AI} {Literacy}},
	url = {https://zenodo.org/records/11613520},
	language = {eng},
	urldate = {2025-03-27},
	publisher = {\#creativeHE.},
	author = {Abegglen, Sandra and Nerantzi, Chrissi and Martínez-Arboleda, Antonio and Karatsiori, Marianna and Atenas, Javiera and Rowell, Chris},
	month = jun,
	year = {2024},
}

@misc{aiartsubreddits,
	author = {{Hive Index}},
	title = {7 {Best} {AI} {Art} {Subreddits} to join in 2025},
	url = {https://thehiveindex.com/topics/ai-art/platform/reddit/},
	language = {en},
	urldate = {2025-03-27},
	year = {2024},
}

@misc{metapix,
	title = {Top 7 {AI}-generated {Art} {Subreddits} for {Creative} {Inspiration}},
	url = {https://medium.com/ai-art-lounge/top-7-ai-generated-art-subreddits-for-creative-inspiration-7e90df0448da},
	language = {en},
	urldate = {2025-03-27},
	journal = {Ai Art Lounge},
	author = {Metapix},
	month = may,
	year = {2024},
}

@misc{nyfa,
	title = {{VISUAL} {ART}-{RELATED} {REDDIT} {SUBREDDITS}},
	url = {https://www.nyfa.org/news/archive/visual-art-related-reddit-subreddits/},
	language = {en-US},
	urldate = {2025-03-27},
	journal = {Nyfa},
	author = {Aronoff, Amy},
	month = aug,
	year = {2015},
}

@misc{format,
	title = {40 {Top} {Subreddits} for {Artists}, {Photographers} and {Designers}},
	url = {https://www.format.com/magazine/resources/art/top-subreddits-art-design/},
	language = {en-US},
	urldate = {2025-03-27},
	journal = {FORMAT},
	author = {Team, Format},
	month = mar,
	year = {2025},
}

@misc{govisually,
	title = {Top 25 subreddits for artists, designers, and photographers},
	url = {https://govisually.com/blog/top-25-subreddits/},
	language = {en-US},
	urldate = {2025-03-27},
	journal = {GoVisually},
	author = {Khan, Alina Zahid},
	month = mar,
	year = {2022},
}

@misc{newartists,
	title = {Art {Subreddits} for {New} and {Veteran} {Artists}},
	url = {https://neoreach.com/art-subreddits/},
	language = {en},
	urldate = {2025-03-27},
	journal = {NeoReach {\textbar} Influencer Marketing Platform},
	author = {Staff, Editorial},
	month = mar,
	year = {2023},
}

@article{aiprof,
	author = {Blit, Joel},
	title = {Opinion: {DeepSeek} just changed the {AI} {Game} — but is {Canada} even playing?},
	shorttitle = {Opinion},
	url = {https://www.theglobeandmail.com/business/commentary/article-deepseek-just-changed-the-ai-game-but-is-canada-even-playing/},
	language = {en-CA},
	urldate = {2025-03-27},
	journal = {The Globe and Mail},
	month = feb,
	year = {2025},
}

@misc{travis,
	title = {{AI} {Rules}? {Characterizing} {Reddit} {Community} {Policies} {Towards} {AI}-{Generated} {Content}},
	shorttitle = {{AI} {Rules}?},
	url = {http://arxiv.org/abs/2410.11698},
	doi = {10.48550/arXiv.2410.11698},
	urldate = {2025-03-25},
	publisher = {arXiv},
	author = {Lloyd, Travis and Gosciak, Jennah and Nguyen, Tung and Naaman, Mor},
	month = mar,
	year = {2025},
}

@misc{prevalence,
	title = {Examining the {Prevalence} and {Dynamics} of {AI}-{Generated} {Media} in {Art} {Subreddits}},
	url = {http://arxiv.org/abs/2410.07302},
	doi = {10.48550/arXiv.2410.07302},
	urldate = {2025-03-03},
	publisher = {arXiv},
	author = {Matatov, Hana and Quéré, Marianne Aubin Le and Amir, Ofra and Naaman, Mor},
	month = oct,
	year = {2024},
}

@misc{midjourney,
	title = {Midjourney},
	url = {https://www.midjourney.com/website},
	urldate = {2025-03-24},
	journal = {Midjourney},
}

@misc{dalle,
	title = {{DALL}·{E}: {Creating} images from text},
	shorttitle = {{DALL}·{E}},
	url = {https://openai.com/index/dall-e/},
	language = {en-US},
	urldate = {2025-03-24},
	month = sep,
	year = {2022},
}

@article{understand1,
	title = {Modeling the structural relationship among primary students’ motivation to learn artificial intelligence},
	volume = {2},
	issn = {2666920X},
	url = {https://linkinghub.elsevier.com/retrieve/pii/S2666920X20300060},
	doi = {10.1016/j.caeai.2020.100006},
	language = {en},
	urldate = {2025-03-23},
	journal = {Computers and Education: Artificial Intelligence},
	author = {Lin, Pei-Yi and Chai, Ching-Sing and Jong, Morris Siu-Yung and Dai, Yun and Guo, Yanmei and Qin, Jianjun},
	year = {2021},
	pages = {100006},
}

@inproceedings{durilong20,
	address = {Honolulu HI USA},
	title = {What is {AI} {Literacy}? {Competencies} and {Design} {Considerations}},
	isbn = {978-1-4503-6708-0},
	shorttitle = {What is {AI} {Literacy}?},
	url = {https://dl.acm.org/doi/10.1145/3313831.3376727},
	doi = {10.1145/3313831.3376727},
	language = {en},
	urldate = {2025-03-18},
	booktitle = {Proceedings of the 2020 {CHI} {Conference} on {Human} {Factors} in {Computing} {Systems}},
	publisher = {ACM},
	author = {Long, Duri and Magerko, Brian},
	month = apr,
	year = {2020},
	pages = {1--16},
}

@article{ng,
	title = {Conceptualizing {AI} literacy: {An} exploratory review},
	volume = {2},
	issn = {2666920X},
	shorttitle = {Conceptualizing {AI} literacy},
	url = {https://linkinghub.elsevier.com/retrieve/pii/S2666920X21000357},
	doi = {10.1016/j.caeai.2021.100041},
	language = {en},
	urldate = {2025-03-22},
	journal = {Computers and Education: Artificial Intelligence},
	author = {Ng, Davy Tsz Kit and Leung, Jac Ka Lok and Chu, Samuel Kai Wah and Qiao, Maggie Shen},
	year = {2021},
	pages = {100041},
}

@article{heyder,
	title = {Extending the foundations of {AI} literacy},
	language = {en},
	author = {Heyder, Teresa and Posegga, Oliver},
	year = {2021},
}

@inproceedings{company,
	address = {IT University of Copenhagen Denmark},
	title = {Generative {AI} in {User} {Experience} {Design} and {Research}: {How} {Do} {UX} {Practitioners}, {Teams}, and {Companies} {Use} {GenAI} in {Industry}?},
	isbn = {9798400705830},
	shorttitle = {Generative {AI} in {User} {Experience} {Design} and {Research}},
	url = {https://dl.acm.org/doi/10.1145/3643834.3660720},
	doi = {10.1145/3643834.3660720},
	language = {en},
	urldate = {2024-12-12},
	booktitle = {Designing {Interactive} {Systems} {Conference}},
	publisher = {ACM},
	author = {Takaffoli, Macy and Li, Sijia and Mäkelä, Ville},
	month = jul,
	year = {2024},
	pages = {1579--1593},
}

@misc{adobe,
	title = {Learn {\textbar} {Adobe}},
	url = {https://creativecloud.adobe.com/cc/learn/app/animate},
	urldate = {2024-09-09},
}

@article{drawing,
	title = {A {Collaborative}, {Interactive} and {Context}-{Aware} {Drawing} {Agent} for {Co}-{Creative} {Design}},
	volume = {30},
	issn = {1941-0506},
	url = {https://ieeexplore.ieee.org/document/10179136/?arnumber=10179136},
	doi = {10.1109/TVCG.2023.3293853},
	number = {8},
	urldate = {2024-09-04},
	journal = {IEEE Transactions on Visualization and Computer Graphics},
	author = {Ibarrola, Francisco and Lawton, Tomas and Grace, Kazjon},
	month = aug,
	year = {2024},
	pages = {5525--5537},
}

@inproceedings{tool,
	address = {Pittsburgh PA USA},
	title = {When is a {Tool} a {Tool}? {User} {Perceptions} of {System} {Agency} in {Human}–{AI} {Co}-{Creative} {Drawing}},
	isbn = {978-1-4503-9893-0},
	shorttitle = {When is a {Tool} a {Tool}?},
	url = {https://dl.acm.org/doi/10.1145/3563657.3595977},
	doi = {10.1145/3563657.3595977},
	language = {en},
	urldate = {2024-09-03},
	booktitle = {Proceedings of the 2023 {ACM} {Designing} {Interactive} {Systems} {Conference}},
	publisher = {ACM},
	author = {Lawton, Tomas and Grace, Kazjon and Ibarrola, Francisco J},
	month = jul,
	year = {2023},
	pages = {1978--1996},
}

@article{project,
	title = {Co-creating with {AI}},
	volume = {2023},
	issn = {0026-5667, 2157-4189},
	url = {https://read.dukeupress.edu/the-minnesota-review/article/2023/100/118/351618/Co-creating-with-AI},
	doi = {10.1215/00265667-10320940},
	language = {en},
	number = {100},
	urldate = {2024-09-03},
	journal = {Minnesota Review},
	author = {Uricchio, William and Cizek, Katerina},
	month = may,
	year = {2023},
	pages = {118--131},
}

@inproceedings{journey,
	address = {Honolulu HI USA},
	title = {Is {It} {AI} or {Is} {It} {Me}? {Understanding} {Users}’ {Prompt} {Journey} with {Text}-to-{Image} {Generative} {AI} {Tools}},
	isbn = {9798400703300},
	shorttitle = {Is {It} {AI} or {Is} {It} {Me}?},
	url = {https://dl.acm.org/doi/10.1145/3613904.3642861},
	doi = {10.1145/3613904.3642861},
	language = {en},
	urldate = {2024-08-12},
	booktitle = {Proceedings of the {CHI} {Conference} on {Human} {Factors} in {Computing} {Systems}},
	publisher = {ACM},
	author = {Mahdavi Goloujeh, Atefeh and Sullivan, Anne and Magerko, Brian},
	month = may,
	year = {2024},
	pages = {1--13},
}

@misc{prompt,
	title = {How to {Write} {Effective} {AI} {Image} {Prompts} [{Examples} + {Tips}]},
	url = {https://dorik.com/blog/how-to-write-ai-image-prompts},
	language = {en},
	urldate = {2024-09-03},
}


\newpage
\appendix
\section{Keyword Taxonomy}
\label{appendix:keyword}

\subsection{Generative AI Platforms and Tools}

This category captures references to widely used commercial and open-source platforms for generating visual content using AI. These tools often act as entry points for artistic experimentation and are frequently mentioned in conversations surrounding creative workflows. Our keyword list includes prominent systems such as Midjourney, Stable Diffusion, ChatGPT, Leonardo AI and ComfyUI.

\begin{itemize}
    \item adobe firefly (firefly)
    \item dalle (dall-e, dalle2, dalle3, dall e)
    \item midjourney (mj)
    \item stable diffusion (stablediffusion, sdxl, stablediff, stablediffusionxl, stable\_diffusion)
    \item deep dream (deepdream, deepdream ai, deepdreamai, deepdream generator, deep dream generator)
    \item runwayml (runway ml)
    \item nightcafe (night cafe)
    \item comfyui (comfy ui)
    \item invokeai (invoke ai)
    \item artbreeder (art breeder)
    \item deepai (deep ai)
    \item deepart (deep art)
    \item starryai (starry ai)
    \item wombo (wombodream, womboo dream)
    \item nightcafe (night cafe)
    \item dreamstudio (dream studio)
    \item leonardo ai (leonardoai)
    \item playgroundai (playground ai)
    \item civitai
    \item chatgpt
    \item deepseek
\end{itemize}

\subsection{AI and Related Concepts}

To reflect varying levels of technical complexity and topical emphasis, we organized terms into a multi-level hierarchy including four different levels, ranging from broadly used AI terms to more advanced technical concepts and ethical considerations (see Table \ref{tab:ai-terminology-levels} for more details).

\begin{table*}[htbp]
  \caption{Taxonomy of AI-related concepts used in keyword filtering. Terms are grouped into four levels reflecting increasing technical complexity and topical depth, ranging from general applications to advanced models and sociotechnical concerns.}
  \label{tab:ai-terminology-levels}  
  \small
  \begin{tabular}{p{5cm} p{9cm}}
    \toprule
    \textbf{\rule{0pt}{2.5ex}Level} & \textbf{\rule{0pt}{2.5ex}Description} \\
    \midrule
    \rule{0pt}{2.5ex}General Terms and Applications & 
      Broad and commonly used terms, such as ai, artificial intelligence and generative ai as well as terms related to user interaction with AI tools, including prompt engineering, prompt crafting, and image prompts. \\[0.3em]
    \rule{0pt}{2.5ex}Core Machine Learning Concepts & 
      Fundamental machine learning concepts are typically encountered in introductory textbooks and courses such as training, dataset, algorithm and loss. \\[0.3em]
    \rule{0pt}{2.5ex}Advanced Models and Architectures & 
      More specialized and technically advanced terms signal a deeper engagement with AI including advanced methods and architectures like generative adversarial networks, transformers, diffusion models and related concepts. \\[0.3em]
    \rule{0pt}{2.5ex}Ethics, Bias, and Explainability & 
      Terms reflecting ethical and sociotechnical considerations in AI, such as ethics, bias, fairness, explainability, and interpretability. \\[0.3em]
    \bottomrule
  \end{tabular}
\end{table*}

\subsubsection{Level 1: General Terms}
\begin{itemize}
    \item ai
    \item artificial intelligence
    \item ai art
    \item ai-generated
    \item ai generation
    \item generative ai
    \item text prompts
    \item prompt design (including prompt crafting, prompt engineering)
    \item ai-assisted
    \item image prompts
    \item prompts
\end{itemize}

\subsubsection{Level 2: Core Machine Learning Concepts}
\begin{itemize}
    \item training
    \item dataset
    \item algorithm
    \item optimization
    \item simulation
    \item data
    \item pattern 
    \item recognition
    \item supervised learning
    \item unsupervised learning
    \item reinforcement learning (rl)
    \item machine learning (machinelearning, ml)
    \item deep learning (deeplearning, dl)
    \item neural network
    \item model
    \item classification
    \item underfitting
    \item overfitting
    \item gradient descent
    \item embedding
    \item loss
    \item recognition
    \item latent space
    \item probabilistic 
    \item reasoning
\end{itemize}

\subsubsection{Level 3: Advanced Models and Architectures}
\begin{itemize}
    \item generative adversarial network (gan)
    \item stylegan
    \item dcgan
    \item cyclegan
    \item pix2pix
    \item conditional gan (cgan and conditionalgan)
    \item transformer (transformer models and transformers)
    \item variational autoencoders (vae)
    \item diffusion models (latent diffusion models)
    \item vqgan (vqgan+clip and vqgan-clip)
    \item discriminator
    \item generator
    \item clip
    \item attention mechanism (cross-attention and cross attention)
\end{itemize}

\subsubsection{Level 4: AI Ethics}
\begin{itemize}
    \item security
    \item ethics
    \item bias
    \item fairness
    \item interoperability
    \item explainability
    \item interpretability
    \item safety
    \item transparency
    \item privacy
\end{itemize}

\subsection{Image Manipulation Techniques}

The final category includes core image generation workflows (e.g. text-to-image and image-to-image) as well as editing operations such as inpainting, upscaling, rendering, and remixing. Also, it includes image-related tasks that involve AI but are not strictly generative, such as image detection and enhancement.

\begin{itemize}
    \item text-to-image (text to image, text2image, txt2img, txt-to-img, txt to img)
    \item image-to-image (image to image, image2image, img2img, img-to-img, img to img)
    \item image generation (image-generation, imagegen, img-gen, image models)
    \item sketch-to-image (sketch to image)
    \item portrait generation
    \item landscape generation
    \item character generation
    \item style transfer
    \item inpainting
    \item outpainting
    \item upscaling
    \item image editing (image manipulation, image enhancement, image restoration)
    \item masking
    \item blending
    \item augmentation
    \item morphological transformations 
    \item colorization
    \item segmentation detection
    \item object detection
    \item rendering
    \item remixing
    \item morphing
    \item transforming
    \item synthesize
\end{itemize}

\section{Subreddit List}
\label{appendix:subreddit}

We initially searched for the term ``AI'' within each subreddit using Reddit's internal search function, which ranks results by relevance. If none of the top five posts contained meaningful references to artificial intelligence, we excluded the subreddit. To ensure the presence of genuine AI-related discourse, we manually reviewed posts surfaced by the ``AI'' search term to verify topical relevance. Subreddits where ``AI'' appeared only tangentially (e.g., unrelated acronyms) were also excluded. The list below is the final list.

\begin{itemize}
  \item abstractart
  \item aianimeart
  \item aiart
  \item alternativeart
  \item amateurart
  \item art
  \item artbuddy
  \item artcrit
  \item artdeco
  \item arthistory
  \item artist
  \item artists
  \item artisthate
  \item artstore
  \item artistlounge
  \item badart
  \item beginner\_art
  \item comicbookart
  \item chatgpt
  \item comic\_crits
  \item conceptart
  \item contemporaryart
  \item canva
  \item creepyart
  \item cryptoart
  \item comics
  \item computergraphics
  \item comfyui
  \item darkgothicart
  \item deepdream
  \item deepseek
  \item dreambooth
  \item deviantart
  \item digitalart
  \item drawforme
  \item drawing
  \item digitalpainting
  \item design
  \item designthought
  \item dalle2
  \item dalle
  \item fantasyart
  \item furryartschool
  \item generative
  \item graphic\_design
  \item hungryartists
  \item idap
  \item illustration
  \item imaginarycharacters
  \item imaginarylandscapes
  \item imaginarymindscapes
  \item imaginarymonsters
  \item imaginarytechnology
  \item imagenai
  \item lightroom
  \item logodesign
  \item midjourney
  \item invokeai
  \item painting
  \item picrequests
  \item pixelart
  \item printmaking
  \item pics
  \item photography
  \item askphotography
  \item promptcraft
  \item postprocessing
  \item sculpture
  \item sketchpad
  \item specart
  \item stencils
  \item streetart
  \item stablediffusion
  \item stablediffusioninfo
  \item sdforall
  \item transformersart
  \item typography
  \item unusualart
  \item watercolor
  \item wombodream
\end{itemize}

\subsubsection{Categorization of the subreddits based on AI Rules}

Through our analysis, we identified four major categories of subreddit rules: (1) \textit{prohibitive rules} that explicitly ban AI-generated content in any form; (2) \textit{conditional allowance rules} that restrict AI discussion under specific conditions; (3) \textit{AI-specific rules} that center around AI art or tools, often with their own unique restrictions; and (4) \textit{implicit rules} that don't explicitly mention AI content.

A number of subreddits have adopted explicit rules banning all AI-generated artwork \cite{travis}. In these spaces, submissions that are partially or entirely produced using AI tools are subject to immediate removal, which may potentially lead to permanent bans. 

Other subreddits take a more moderated approach, permitting AI-related content under specific conditions. These restrictions often include posting quotas (e.g., one AI-generated post per user per week), content flair requirements or limitations on specific types of AI-related discussion (e.g. no AI art is art /not art debate or posts must discuss topics that haven't been recently or frequently discussed). Two subreddits (r/painting and r/drawing) in this category do not explicitly ban AI posts but prohibit low-effort questions like ``What style is this?'' or ``Is this good for X?'', which are often seen as prompts used to train AI models. 

In contrast to subreddits that conditionally permit AI-generated content, \textit{r/aiArt}---a community centered on AI-generated artwork—enforces AI-specific guidelines that explicitly prohibit users from questioning whether an image was created with AI tools. Other communities adopt even more stringent approaches. For instance, \textit{r/aianimeart} mandates that all submissions be AI-generated, making such content a prerequisite for participation. Likewise, \textit{r/midjourney} restricts submissions exclusively to images produced using the Midjourney platform, embedding tool-specific creation as a core normative expectation.

Finally, some subreddits do not explicitly address AI-generated content in their posting guidelines.

\section{Topic Modeling Results}
\label{appendix:topic_modeling}

We trained models with topic counts \(K\) ranging from 20 to 40 in increments of 5, setting Dirichlet priors \(\alpha = 0.1\) and \(\eta = 0.01\), and ran 2,000 Gibbs-sampling iterations for each configuration. We also extracted up to 500 n-grams (unigrams to trigrams) that appear at least 20 times in the corpus and at least 10 documents using Tomotopy's built-in method. We removed stopwords by combining Gensim's standard stopword list \cite{gensim} with a custom set of additional terms, yielding a final list of 380 stopwords. Using this list, we computed topic model coherence scores using UMass \cite{umass}, C\_v \cite{cvmeasure} and NPMI \cite{npmi} metrics.

\begin{table}[htbp]
    \centering
    \caption{Comparison of Coherence Scores Across Different Topic Numbers For Posts}
    \small
    \begin{tabular}{cccc}
        \toprule
        \textbf{Num. Topics} & \textbf{C\_v} & \textbf{U\_Mass} & \textbf{C\_NPMI} \\
        \midrule
        20 & 0.61 & -2.51 & 0.06 \\
        25 & 0.62 & -2.56 & 0.07 \\
        30 & 0.62 & -2.63 & 0.07 \\
        35 & 0.62 & -2.55 & 0.07 \\
        40 & 0.62 & -2.64 & 0.07 \\
        \bottomrule
    \end{tabular}
    \label{tab:posts_coherence_scores}
\end{table}

\begin{table}[htbp]
    \centering
    \caption{Comparison of Coherence Scores Across Different Topic Numbers For Comments}
    \small
    \begin{tabular}{cccc}
        \toprule
        \textbf{Num. Topics} & \textbf{C\_v} & \textbf{U\_Mass} & \textbf{C\_NPMI} \\
        \midrule
        20 & 0.69 & -2.34 & 0.05 \\
        25 & 0.68 & -2.41 & 0.06 \\
        30 & 0.67 & -2.44 & 0.06 \\
        35 & 0.68 & -2.43 & 0.06 \\
        40 & 0.67 & -2.63 & 0.06 \\
        \bottomrule
    \end{tabular}
    \label{tab:comments_coherence_scores}
\end{table}

\renewcommand{\arraystretch}{1.2} 

\begin{table*}[htbp]
\centering
\caption{Topics discovered by topic modeling, their probabilities in the dataset, and representative tokens. The percentage of each topic was calculated based on the number of posts assigned to that topic in the whole dataset. This list included 24 topics derived from posts, and we then discarded those not related to our research questions (labeled as unrelated), resulting in 18 topics included in the subsequent analysis.}
\label{tab:post_topics}
\small
\renewcommand{\arraystretch}{1.1}
\setlength{\tabcolsep}{6pt}
\begin{tabular}{p{5cm} p{1.5cm} p{9cm}}
\toprule
\textbf{Topic} & \textbf{Prob.} & \textbf{Representative Tokens} \\
\midrule
Basic Setup \& Getting Started Help & 23.8\% & use, help, ai, looking, way, trying, create, good, id, model \\
Practical Tool Failures & 15.6\% & tried, got, didnt, work, try, started, problem, issue, trying, getting \\
Output Quality Complaints \& Policy Concerns & 10.6\% & people, ai, think, good, way, better, right, work, feel, real \\
Model Training \& Fine-Tuning & 7.2\% & model, image, images, models, lora, training, use, sd, sdxl, results \\
Prompt Vocabulary & 5.6\% & image, prompt, style, character, midjourney, background, create, dalle, example, color \\
AI Visual Art Sharing \& Community Promotion & 4.5\% & ai, art, artists, post, image, midjourney, share, community, free, artists \\
ChatGPT Limitations \& Performance Issues & 3.7\% & chatgpt, gpt, text, prompt, write, use, ask, code, words, chat \\
Artwork and Jobs (unrelated) & 3.6\% & art, design, work, drawing, job, years, artist, graphic, digital, draw \\
ChatGPT Access \& Subscription Issues (unrelated) & 3.5\% & chatgpt, gpt, app, api, chat, use, access, openai, free, account \\
ComfyUI \& Node Workflow Problems & 3.0\% & file, image, use, prompt, comfyui, node, folder, code, workflow, add \\
Lay Explanations of AI Mechanisms & 2.7\% & model, data, number, based, different, example, process, given, specific, case \\
AI Tool Integration \& Community Collaboration & 2.6\% & ai, models, data, tools, new, tool, users, project, content, feedback \\
ChatGPT Jailbreaks \& Persona Roleplay & 1.7\% & chatgpt, information, provide, dan, prompt, answer, user, response, ask \\
AI Industry News \& Public Discourse & 1.6\% & ai, openai, data, company, use, new, content, public, technology, legal \\
Photography (unrelated) & 1.5\% & photos, camera, photography, photo, lightroom, lens, light, shoot, editing, raw \\
AI Capabilities \& Ethical Reflections & 1.5\% & ai, human, world, understanding, potential, systems, language, life, ethical \\
Configuration \& Setup & 1.2\% & stable\_diffusion, pc, free, gpu, run, vram, automatic1111\_web\_ui, sd, ram, colab \\
Installation \& Runtime Errors & 1.2\% & error, install, file, version, python, cude, line, installed, torch, xformers \\
Prompts Parameter Settings & 1.1\% & prompt, quality, negative, model, bad, seed, detailed, size, steps, hands \\
Narrative Outputs Sharing (unrelated) & 1.1\% & world, life, journey, light, power, universe, new, space, earth, dark \\
Model Runtime Errors \& Debugging & 0.7\% & file, line, return, error, false, import, shape, model, kwargs, traceback\_most\_recent \\
ChatGPT Game Play \& Role Interaction (unrelated) & 0.4\% & game, player, games, day, play, players, water, end, chatgpt, character \\
Political Narratives \& Debates with ChatGPT (unrelated) & 0.3\% & war, political, country, government, president, trump, rights, military, nuclear, american \\
Code Snippets \& Debugging Help & 0.3\% & false, true, import, e, c, n, def, type, return, x \\
\bottomrule
\end{tabular}
\end{table*}

\begin{table*}[htbp]
  \centering
  \caption{The topics discovered by topic modeling of comments and the top 10 tokens associated with each topic. The percentage of each topic was calculated by the number of comments assigned to that topic in the whole dataset. Each comment was assigned to its most probable topic, and each topic was initially characterized by its top-N words \cite{blei}.}
  \label{tab:comments_topic}
  \small
  \renewcommand{\arraystretch}{1.1}
  \setlength{\tabcolsep}{6pt}
  \begin{tabular}{p{5cm} p{1.5cm} p{9cm}}
    \toprule
    \textbf{Topic} & \textbf{Percentage} & \textbf{Representative Tokens} \\
    \midrule
    ChatGPT Behavior, Failures, and Jailbreak Workarounds & 14.4\% & ai, think, good, chatgpt, got, people, work, didnt, prompts, love \\
    Image Generation Prompts, Models, and Techniques & 12.5\% & model, image, images, use, prompt, models, training, prompts, sd, sdxl \\
    Copyright, Fair Use, and Artist Rights & 9.1\% & ai, art, people, artists, artist, work, images, use, think, image \\
    AI Industry, Hype, and Labor Impacts & 8.7\% & ai, people, work, think, money, years, good, job, use, new \\
    Model Capabilities, Limits, and Misconceptions & 7.7\% & ai, chatgpt, data, think, human, way, model, people, training, humans \\
    Automated Moderation Notices & 6.7\% & post, questions, concerns, moderators, bot, automatically\_please\_contact, subredditmessagecomposeto, comment, group, sharing \\
    ChatGPT API Usage and Access Issues & 6.6\% & chatgpt, use, gpt, prompt, code, ask, ai, data, model, openai \\
    Errors, Crashes, and Debugging Help & 4.5\% & model, use, comfyui, file, run, vram, models, stable\_diffusion, gpu, sdxl \\
    Bot Rule Enforcement (ChatGPT / DALL·E) & 3.6\% & image, post, prompt, reply, conversation, concerns, chatgpt, bots, dalle, consider\_joining \\
    Automod Link \& Source Rules & 3.2\% & links, use, share, external, source, rules, post, add, correct, real \\
    General Bot Removal Notices & 2.8\% & bot, prompt, comment, questions, chatgpt, moderators, concerns, automatically\_please\_contact, subredditmessagecomposeto, public\_discord\_server \\
    Bot Flags for AI Model Mentions & 2.3\% & bot, ai, prompt, chatgpt, post, model, opensource, perplexity, open\_assistant, generator \\
    Photography Techniques and Camera Gear & 2.2\% & camera, use, photos, lens, model, photography, photo, good, data, image \\
    How Language Models Work & 1.1\% & ai, human, language, information, data, provide, content, use, chatgpt, potential \\
    Prompt Parameters Settings & 1.1\% & prompt, model, quality, steps, detailed, bad, negative, cfg, scale, extra \\
    Artists and Jobs & 0.4\% & artist, check, conversation, report, comment, search, artists, username, reviews, client \\
    \bottomrule
  \end{tabular}
\end{table*}

\begin{table*}[htbp]
\centering
\caption{Overview of the six themes distilled from topic modeling, ranked by prevalence in descending order. Each theme captures a distinct mode of engagement with AI tools among creators, reflecting its relative frequency in the dataset.}
\label{tab:merged_themes}
\small
\setlength\tabcolsep{12pt} 
\begin{tabular}{@{} p{4cm} c p{8cm} @{}} 
\toprule
\textbf{Theme} & \textbf{Percentage} & \textbf{Details}\\
\midrule
Tool Related Complaints & 26.2\% & 
Merged from Practical Tool Failures (15.6\%) and Output Quality Complaints \& Policy Concerns (10.6\%), capturing frustrations with broken tools, poor results, and restrictive policies. \\[0.6em] 

Basic Setup \& Getting Started Help & 23.8\% & 
Original theme with no merging required due to substantial representation. It captures beginner-oriented posts where users ask for basic guidance, often phrased as ``can someone help me...'' or ``how do I...'', about setting up, choosing, or using generative AI tools.\\[0.6em] 

Model Training \& Workflow Customization & 13.3\% & 
Merged from Model Training \& Fine-Tuning (7.2\%), ComfyUI \& Node Workflow Problems (3.0\%), Configuration \& Setup (1.2\%), Installation \& Runtime Errors (1.2\%), and Model Runtime Errors \& Debugging (0.7\%), covering advanced customization of models and technical troubleshooting. \\[0.6em] 

Broader AI Reflections & 11.2\% & 
Merged from AI Industry News \& Public Discourse (1.6\%), AI Capabilities \& Ethical Reflections (1.5\%), Lay Explanations of AI Mechanisms (2.7\%), ChatGPT Limitations \& Performance Issues (3.7\%), and ChatGPT Jailbreaks \& Persona Roleplay (1.7\%), covering ethical debates, industry news, lay sensemaking of AI, and user reflections on ChatGPT's limitations and jailbreak practices. \\[0.3em] 

Sharing, Feedback \& Community & 7.1\% & 
Merged from AI Art Sharing \& Community Engagement (4.5\%) and AI Tool Integration \& Community Collaboration (2.6\%), highlighting social practices of sharing outputs, critique, and collaboration. \\[0.6em] 

Prompting Practices \& Refinement & 6.7\% & 
Merged from Prompt Vocabulary (5.6\%) and Prompt Parameter Settings Techniques (1.1\%), reflecting how users refine prompt vocabulary and prompt parameters to improve AI outputs. \\[0.6em] 

\bottomrule
\end{tabular}
\end{table*}

\newpage
\section{Codebook}
\label{appendix:codebook}

\textit{Note: This codebook treats tools and models as the same, because novices may not be able to distinguish between them when discussing related topics.}

\subsection {Tool Literacy}
Captures how users understand and manipulate AI tools' features, configurations, and troubleshooting practices.

\begin{description}
    \item[\textbf{Hardware Configuration}]  
    When users discuss the computing environment or hardware requirements for running AI tools.
    \item[\textbf{Access \& Authentication}]  
    When users discuss account access methods, subscription tiers, API keys, rate limits, or billing.
    \item[\textbf{Help Seeking}]
    When users explicitly ask peers for assistance with debugging errors, procedural questions, workflow queries, or tool recommendations.
    \begin{description}
        \item[\textbf{Procedure Help}] 
        When users post about steps taken to finish a specific task, or are confused about where to start as their first step. Typical posts may be like ``How do I do this with [the tool]?'' or ``What do I do first?''. 
        
        \item[\textbf{Interpretive Questions}] 
        When users look for an explanation for the problem they encounter. For example, typical posts may be ``Why does this happen?'' or  ``What did I do wrong?''. 

        \item[\textbf{Determine Possibilities}] 
        When users are not sure about if it is possible to do the task they want with certain tools or if there is any tools that can do the task they want. Posts might be ``Can I do this with the...[tool]?'', ``Is there any application that can do this?''.
        
        \item[\textbf{Descriptive Questions}] 
        When users seek help from others to describe something rather than asking why it happens or how to fix it. For example, ``What is this?'' or ``What is the difference between...?''.
        
        \item[\textbf{Troubleshooting}]
        When users need help debugging the tool or changing its settings after a process fails, such as hyperparameters, they ask for assistance. For example, ``How do I fix it?''.

        \item[\textbf{Resource Recommendation Requests}]
        When users ask for external assets or guidance materials needed to work with a tool, rather than asking how to do something or whether it's possible, they're asking where to get what they need or if someone can share it. For example, ``Are there tutorials about...?''. 

        \item[\textbf{Prompt Feedback}]
        When users have their prompts in the post, they are looking for feedback for improving the results or seeking suggestions, ideas for revising the prompts, or asking for assistance writing prompts. This differs from prompt sharing, where the goal is to showcase a prompt example rather than request input or critique.

    \end{description}
    \item[\textbf{Tool Comparison}] 
    When users compare the functionality of different tools or different models of the same tool.

    \item[\textbf{Prompt Sharing}]
    When users post their full prompt as an illustrative example of prompt-crafting, with or without sharing their negative prompting.

\end{description}

\subsection {Capacity Awareness}
Capture users' reflections on model behavior, including biases, limitations, and capability exploration.

\begin{description}
    \item[\textbf{Limitations Recognition}]  
    When they call out model failures.
    \item[\textbf{Capability Exploration}] 
    When users pose challenging or boundary-pushing prompts to explore what the model is capable of.
    
    \begin{description}
        \item[\textbf{Being Curious}]
        When users explore the capacity of the tools because of their curiosity about the output.
        \item[\textbf{Explore Capacity}] 
        When users are exploring whether the model is capable of doing specific tasks. 
    \end{description}
    
    \item[\textbf{Internal Mechanism}]  
    When users describe or ask about the internal workings or algorithms of the model.

    \item[\textbf{Strength Recognition}]  
    When users reflect on how they apply AI tools in specific contexts, they share what has worked well for them.

\end{description}

\subsection {Ethics and Responsible Use}
Capture users' conversations about the ethical, legal, and safety implications of AI use.

\begin{description}
    \item[\textbf{Comparing AI with Humans}]  
    When users compare AI’s behavior to human behavior (e.g., whether AI can be ``caring'', ``fair'', or ``empathetic'' like humans). 
    \item[\textbf{Bias}]  
    When users suspect bias in AI output or the training dataset.
    \item[\textbf{Copyrights Concerns}]  
    When they worry about copyright or ownership of AI outputs.  
    \item[\textbf{Data Privacy}]  
    When users express concerns about data privacy issues in the training dataset or personal data usage. 
    
    \item[\textbf{Misuse \& Safety}]  
    When users criticize an AI model's safety measures---arguing that safeguards degrade output quality---and users who deliberately try to bypass those protections to generate disallowed, sensitive, or malicious content. It also covers conversations of model misuse. 
    
    \item[\textbf{Impact of AI}]
    Discussion about AI's impact on multiple aspects, including impact on jobs, education and creative work, etc. 
    
    \item[\textbf{AI Lab Policy}]
    When users discuss terms or conditions associated with AI tools/companies.

\end{description} 

\subsection{Community Engagement}
Captures how users engage with workflow sharing, giving feedback, and sharing resources within the community.

\begin{description}
    \item[\textbf{Workflow Sharing}]  
    When users post their workflow with or without outputs for others.  
    \item[\textbf{Peer Feedback}]  
    Seeking or offering critiques on outputs or projects.    
    \item[\textbf{Resource Sharing}]
    When users share resources they've created or found helpful, such as tools, models, tutorials, or blogs.
\end{description}

\subsection {Promotion}
When users engage in promotional activities like promoting their own projects, directing attention to external platforms (e.g., social media channels, portfolios, or commercial websites). Posts in this category are primarily intended to increase visibility, attract audiences, or generate professional opportunities, rather than to seek feedback or exchange knowledge.

\subsection{AI Output Sharing}
When users share the output made using AI tools, the goal of the post is to showcase their work rather than seek feedback to improve it.

\subsection{AI Tech Dynamics Sharing}
When users share model release news, AI tech company news, or blog posts related to AI technologies.

\subsection{Not Related Content} When users post not AI-related content, like photography, jobs, commissions, or content not relevant to our research questions.

\section{Classification Approaches}
\label{appendix:classification_methods}
\subsection{Regular Expressions}

Our initial rule-based approach involved manually writing seven regular expressions---one per class (excluding ``Not Related Content'', which served as a fallback category). This approach yielded only 32.5\% accuracy, as regex patterns proved too limited and failed to capture the nuanced semantics of user language.

\subsection{SVM with TF-IDF features}

Next, we trained a linear SVM classifier using TF-IDF features. With a training set of 600 and test set of 300, we achieved 45\% accuracy---an improvement, though still limited by the model's inability to generalize across semantic boundaries.

\subsection{LLM - Gemini 2.5 Flash}

We then explored prompting Gemini 2.5 Flash with 12 manually curated examples, using a zero-shot-like strategy. These examples represented the four core literacy categories---Tool Literacy, Community Engagement, Capacity Awareness, and Ethics and Responsible Use, with all remaining classes grouped into a fallback category. This approach reached 64\% accuracy on the remaining 888 conversations. We observed that smaller batch sizes (5 conversations per request) consistently outperformed larger ones, likely due to context window limitations. However, Gemini's content moderation policies caused it to silently fail on posts containing profanity or NSFW content, limiting its usability.

\subsection{Fine-Tuned BERT}

As a next step, we experimented with fine-tuning a BERT model, but early results were highly ineffective. The model struggled with data sparsity and class imbalance, achieving only 20\% accuracy and mostly predicting the Tool Literacy category. Given this poor performance and the high computational cost, we chose not to pursue this approach further.

\subsection{LLM - Claude Haiku 3}

Our final and most successful method involved using the Claude API. We began with Claude Haiku 3 and a custom prompt that defined all 8 classes (provided in the Appendix~\ref{appendix:prompt}). We experimented with a retrieval-augmented generation (RAG) approach using the all-MiniLM-L6-v2 model from Sentence Transformers. A vector database of 600 manually labeled conversations was created. At inference time, up to 5 semantically similar examples (by cosine similarity) were retrieved and included in the prompt to assist the model's prediction. However, this RAG pipeline offered minimal improvement, so it was ultimately not used in our final approach.

\section{LLM Prompt for Classification}
\label{appendix:prompt}

The following prompt is used for classification of the Reddit conversations. Note that the placeholders \{\{post\}\} and \{\{examples\}\} are dynamically instantiated for each classification task. Specifically, {{post}} contains the content of the Reddit post to classify, while {{examples}} is populated with contextually relevant example posts and their associated labels, retrieved dynamically to assist the LLM in accurate categorization.

\begin{lstlisting}[style=promptstyle]
    You are an expert at classifying Reddit posts about AI creativity and literacy.
    Classify the following post into one of these categories:
    <categories>
    <category>
        <label>Tool Literacy</label>
        <content>Posts specifically about HOW to use AI tools and technical implementation
        - Step-by-step tutorials, troubleshooting guides
        - Prompt engineering techniques, prompt sharing
        - API usage, authentication, billing questions
        - Software/hardware requirements, installation help
        - Tool comparisons focused on functionality
        - "How do I..." or "Help with..." posts
        </content>
    </category>
    <category>
        <label>Capacity Awareness</label>
        <content>Posts about WHAT AI can and cannot do (capabilities/limitations)
        - Testing AI model capabilities or limitations
        - "Can AI do X?" questions
        - Discussing model performance, accuracy, failures
        - Understanding how AI models work internally
        - Benchmarking, comparing AI vs human performance
        - Posts exploring AI boundaries and possibilities
        </content>
    </category>
    <category>
        <label>Ethics and responsible use</label>
        <content>Posts about moral, legal, and responsible AI usage
        - Bias, fairness, discrimination concerns
        - Privacy, copyright, data protection issues
        - Safety, misinformation, harmful applications
        - Impact on jobs, education, society
        - Terms of service, legal compliance
        - Jailbreaking, bypassing safety measures
        </content>
    </category>
    <category>
        <label>Community Engagement</label>
        <content>Posts that seek community interaction and feedback
        - "What do you think about..." or "Rate my..." posts
        - Seeking advice, opinions, or recommendations
        - Asking for project feedback or collaboration
        - Discussion starters, polls, surveys
        - Sharing workflows seeking input
        </content>
    </category>
    <category>
        <label>AI Output Sharing</label>
        <content>Posts showcasing AI-generated content
        - Images, text, code, music created with AI
        - "Look what I made with AI" posts
        - Before/after comparisons of AI outputs
        - Sharing results from AI tools (DALL-E, ChatGPT, etc.)
        - Creative experiments with AI
        </content>
    </category>
    <category>
        <label>Promotion</label>
        <content>Self-promotional posts advertising services or products
        - AI consulting services, freelance offerings
        - Tool releases, product launches
        - Personal projects seeking clients/users
        - Marketing content, advertisements
        - "I built this..." with commercial intent
        </content>
    </category>
    <category>
        <label>AI Tech Dynamics Sharing</label>
        <content>Posts about AI industry news and developments
        - New model releases, updates
        - Research paper discussions
        - Company announcements, policy changes
        - Industry trends, market analysis
        - Technical breakthroughs, innovations
        </content>
    </category>
    <category>
        <label>Not related content</label>
        <content>Everything else. Posts not substantially related to AI
        - General programming questions
        - Non-AI creative work
        - Off-topic discussions
        - Spam or irrelevant content
        </content>
    </category>
    </categories>
    
    Here is the reddit post:
    <post>
        {{post}}
    </post>
    
    Use the following examples to help you classify the query:
    <examples>
        {{examples}}
    </examples>
    
    Respond with just the label of the category between category tags. Use the 'Not related content' category as a last resort.
\end{lstlisting}

The effectiveness of our final classification approach - LLM-based with Claude Sonnet 3.5---varied across categories: ``Tool Literacy'' achieves the highest F1-score of 89\%, ``Ethics and Responsible Use'' also performs great with an F1-score of 88\%. ``Community Engagement'' shows more mixed results with a moderate F1-score of 0.67 and a lower recall of 62\%, suggesting that the system may miss some nuanced community discussions that weren't as well represented in the training posts. ``Capacity Awareness'' presents a trade-off with very high precision (93\%) but a significantly lower recall (64\%). The remaining categories show varied performance, with ``Promotion'' achieving perfect balance and others having different precision-recall trade-offs based on how well their patterns were captured in the prompt and the training data.

\section{Plots for Tool-Specific Subreddits and General Creative Subreddits}
\label{appendix:plots}

To examine whether the tool-dominance pattern might be shaped by subreddit selection bias, we separated the dataset into general creative subreddits and tool-specific subreddits. The tool-specific group included r/midjourney, r/dalle, r/dalle2, r/stablediffusion, r/stablediffusioninfo, r/sdforall, r/comfyui, r/invokeai, r/dreambooth, r/deepdream, r/deepseek, r/chatgpt, and r/wombodream. We then added one additional analysis on theme patterns (including raw counts and percentages) for each group and the plots are listed below.

\begin{figure*}[htpb]
  \centering
  \includegraphics[width=1.0\linewidth]{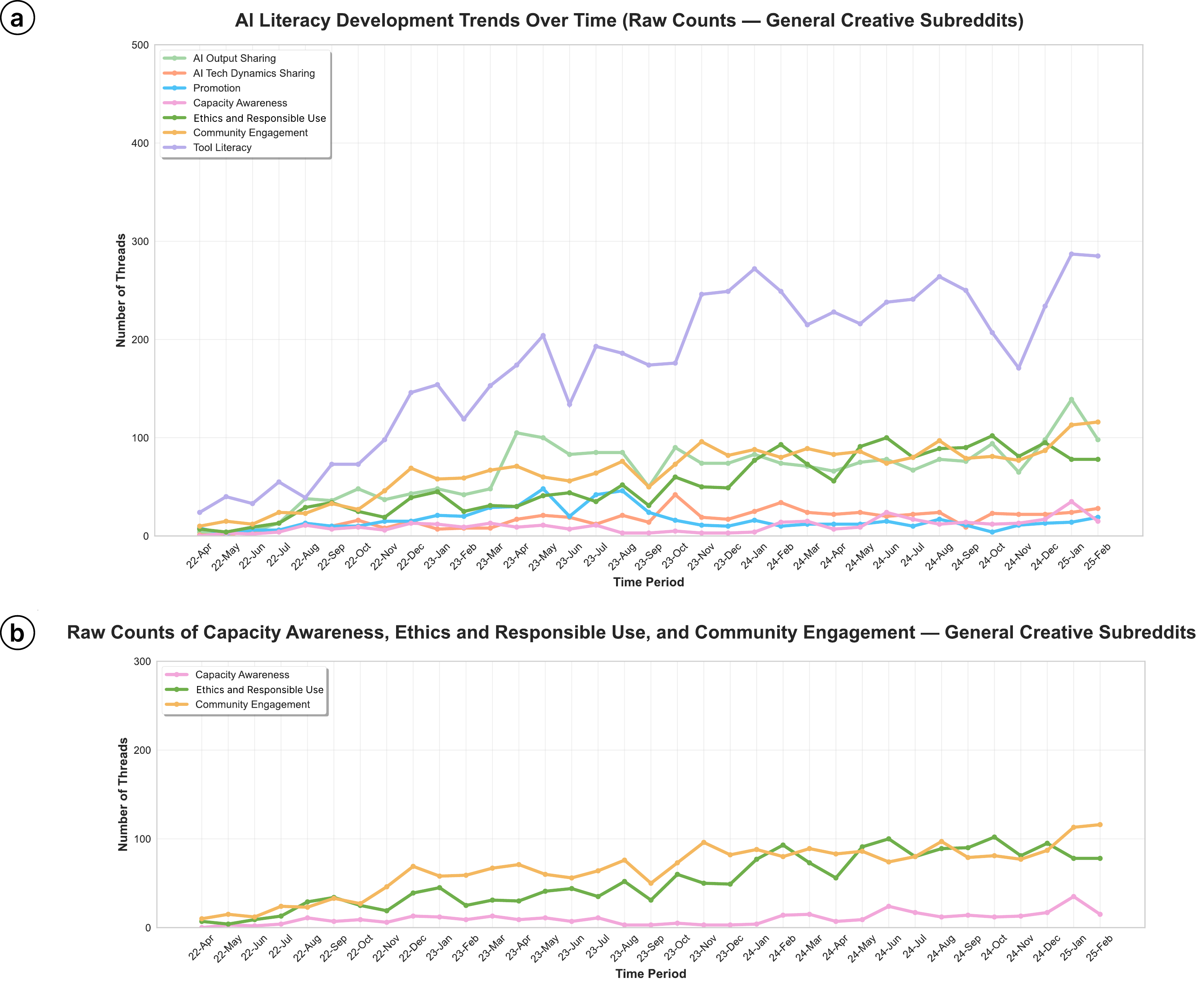}
  \caption{Panel (a) presents the raw thread counts for each AI literacy theme across time, revealing growth patterns and fluctuations in posting volume within general creative communities. Panel (b) highlights raw counts for Capacity Awareness, Ethics and Responsible Use, and Community Engagement. Compared to tool-specific subreddits, general creative communities generate far fewer threads overall, which results in lower raw counts across all AI literacy themes. \textit{Please note that subcharts (a) and (b) are using different y-axis scales.}}
  \label{fig:generalraw}
  \Description{The figure shows raw counts of AI literacy discussions over time within general creative subreddits. Panel a presents thread counts across all AI literacy themes, showing overall growth and fluctuations, with Tool Literacy consistently having the highest volume and other themes appearing at lower levels. Panel b isolates Capacity Awareness, Ethics and Responsible Use, and Community Engagement using a different vertical scale, highlighting gradual increases and periodic variation. Overall, the figure shows that general creative subreddits have lower discussion volume than tool specific communities, but still exhibit steady engagement with multiple AI literacy themes over time.}
\end{figure*}

\begin{figure*}[tb]
  \centering
  \includegraphics[width=1.0\linewidth]{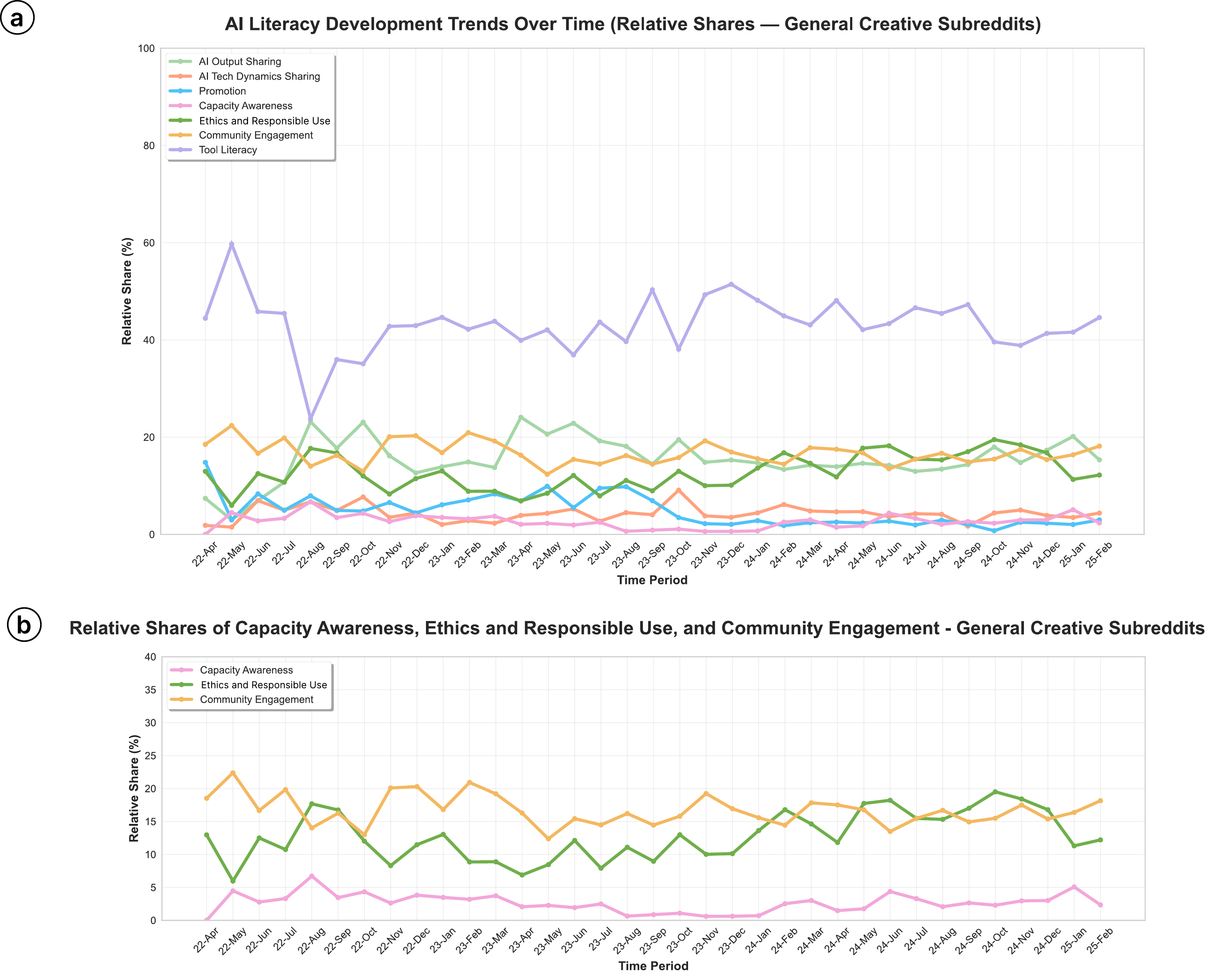}
  \caption{Panel (a) shows the relative shares of six AI literacy themes over time for general creative communities. Tool Literacy remains the largest category but shares the space with Capacity Awareness, Community Engagement, and Promotion, producing a more balanced distribution of themes. Panel (b) focuses on three less frequent but important themes and visualizes their relative shares over time. Together, these trends show that general creative subreddits engage with a broad mix of AI literacy dimensions rather than concentrating primarily on tool use. \textit{Please note that subcharts (a) and (b) are using different y-axis scales.}}
  \label{fig:generalpercentage}
  \Description{The figure shows relative shares of AI literacy discussions over time within general creative subreddits. Panel a presents the proportion of conversations across themes, with Tool Literacy remaining the largest share but alongside substantial contributions from Capacity Awareness, Community Engagement, Promotion, and other themes, resulting in a more balanced distribution than in tool focused communities. Panel b highlights Capacity Awareness, Ethics and Responsible Use, and Community Engagement using a different vertical scale, showing moderate fluctuations and sustained presence over time. Together, the panels indicate that general creative subreddits engage with a broad mix of AI literacy dimensions rather than focusing primarily on tool use.}
\end{figure*}

\begin{figure*}[tb]
  \centering
  \includegraphics[width=1.0\linewidth]{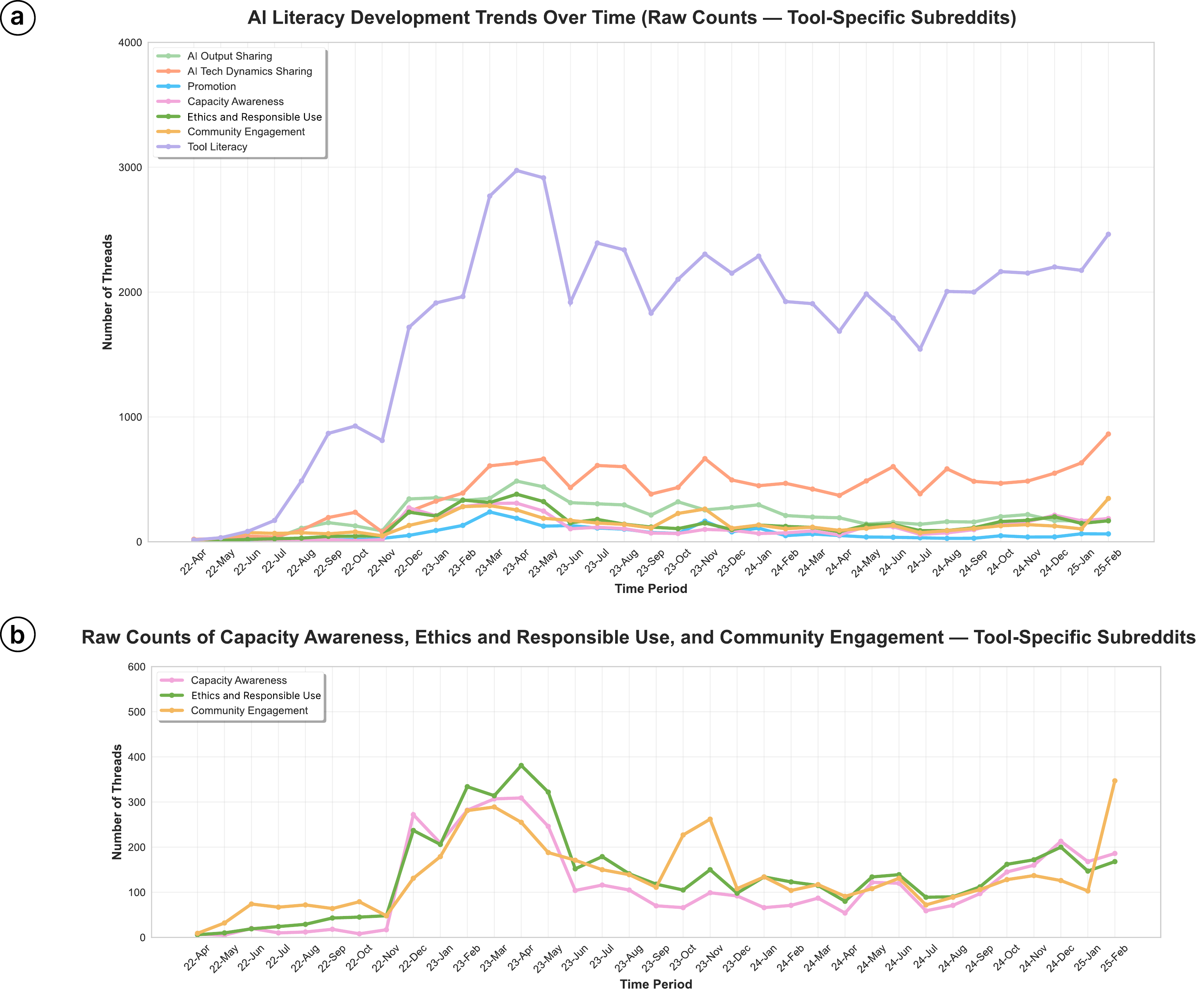}
  \caption{Panel (a) shows the raw conversation counts for all AI literacy-related themes, illustrating substantially higher posting volume in tool-specific subreddits and the strong prominence of Tool Literacy threads. Panel (b) shows raw counts for Capacity Awareness, Ethics and Responsible Use, and Community Engagement. \textit{Please note that subcharts (a) and (b) are using different y-axis scales.}}
  \label{fig:topicraw}
  \Description{The figure shows raw counts of AI literacy discussions over time within tool specific subreddits. Panel a presents thread counts across all themes, with Tool Literacy dominating discussion volume and showing large spikes over time, while other themes such as AI Output Sharing, Capacity Awareness, Ethics and Responsible Use, and Community Engagement appear at much lower levels. Panel b isolates Capacity Awareness, Ethics and Responsible Use, and Community Engagement using a different vertical scale, showing episodic increases around major AI developments. Overall, the figure highlights the substantially higher posting volume and strong tool centered focus of tool specific communities.}
\end{figure*}

\begin{figure*}[tb]
  \centering
  \includegraphics[width=1.0\linewidth]{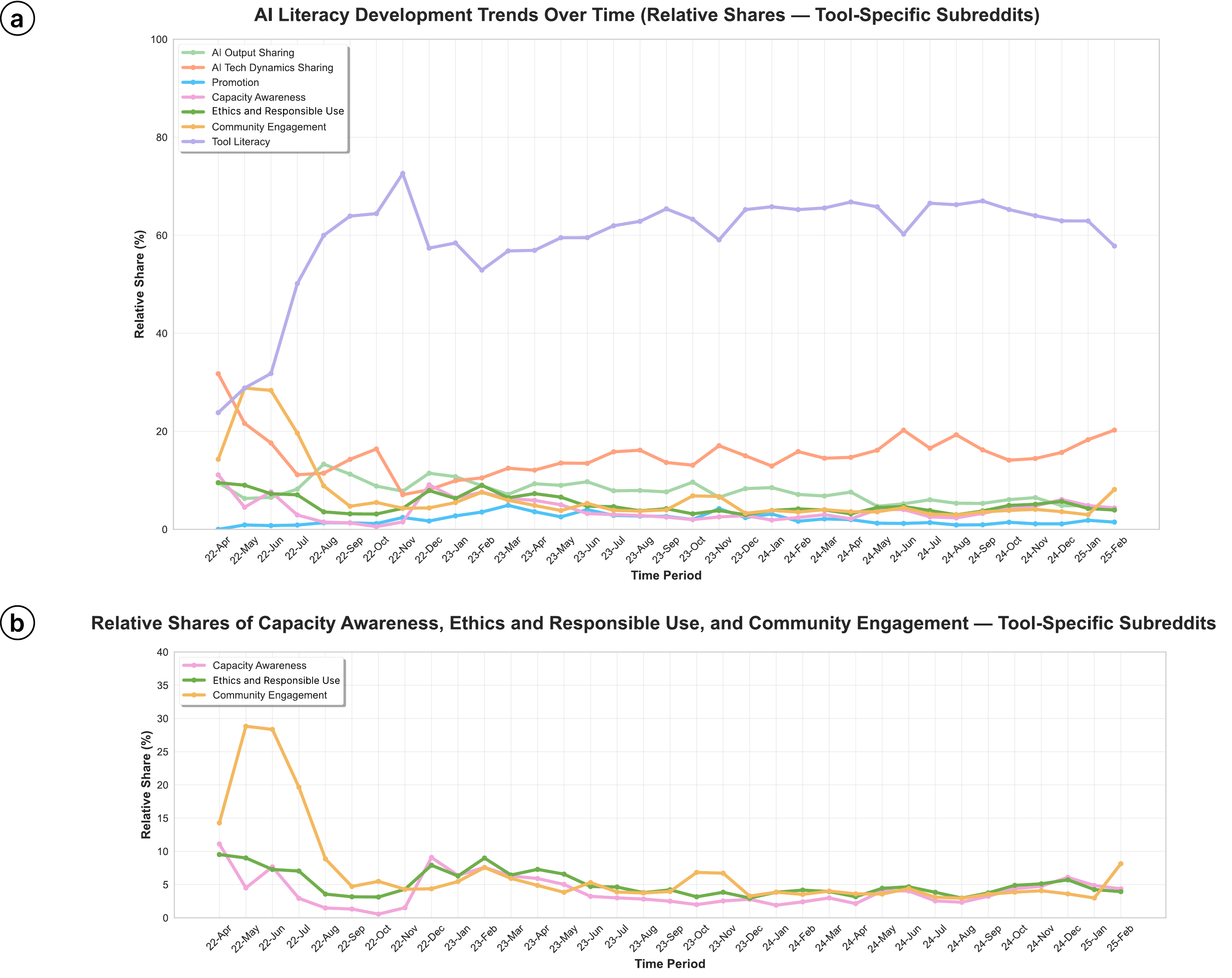}
  \caption{The top panel (a) shows the relative share of AI-literacy conversations within tool-specific subreddits over time. The bottom panel (b) isolates trends in Capacity Awareness, Ethics and Responsible Use, and Community Engagement, illustrating how these themes evolve within communities where Tool Literacy overwhelmingly dominates day-to-day discussion. \textit{Please note that subcharts (a) and (b) are using different y-axis scales.}}
  \label{fig:topicpercentage}
  \Description{The figure shows relative shares of AI literacy discussions over time within tool specific subreddits. Panel a presents the proportion of conversations across themes, with Tool Literacy consistently accounting for the majority of discussion, while themes such as Capacity Awareness, Ethics and Responsible Use, Community Engagement, and AI Output Sharing occupy much smaller shares. Panel b focuses on Capacity Awareness, Ethics and Responsible Use, and Community Engagement using a different vertical scale, showing brief fluctuations but sustained low relative presence. Overall, the figure highlights the strong dominance of tool focused discussion within tool specific communities.}
\end{figure*}

\end{document}